\documentclass[12pts,preprint]{aastex}

\usepackage{lscape}
\usepackage{graphics}
\usepackage{graphicx}
\usepackage{natbib}

\shorttitle{Super-Earth in stellar field} \shortauthors{Laine and Lin}

\begin{document}
\bibliographystyle{biblatex}
\title{Interaction of Close-in Planets with the Magnetosphere of their
Host Stars. II. Super-Earths as Unipolar Inductors and their Orbital Evolution}

\author{Randy O. Laine$^{1,2}$ and Douglas N.C. Lin$^{3,4}$ }

\affil{$^1$ Ecole Normale Superieure, Paris, France, randy.laine@normalesup.org}
\affil{$^2$ LPC2E, Universite d'Orleans/CNRS, Orleans, France, randy.laine@cnrs-orleans.fr } 
\affil{$^3$Department of Astronomy and Astrophysics, University of California, Santa Cruz, CA 95064, USA,
lin@ucolick.org}
\affil{$^4$Kavli Institute of Astronomy \&
Astrophysics, Peking University, Beijing, China}

\begin{abstract} 
Planets with several Earth masses and a few day orbital periods 
have been discovered through radial velocity and transit 
surveys. Regardless of their formation mechanism, an important evolution 
issue is the efficiency of their retention in the proximity of their 
host stars. If these ``super-Earths'' attained their present-day orbits 
during or shortly after the T Tauri phase of their host stars, a large 
fraction of these planets would have encountered intense stellar magnetic field. 
Since these rocky planets have a higher conductivity than the 
atmosphere of their host stars, the magnetic flux tube connecting 
them would slip though the envelope of the host stars faster 
than across the planets. The induced electro-motive force 
across the planet's diameter leads to a potential drop which 
propagates along a flux tube away from the planet with an Alfven speed.  
The foot of the flux tube would sweep across the stellar surface and
the potential drop across the field lines drives a DC current 
analogous to that proposed for the electro-dynamics of the Io-Jupiter
system. The ohmic dissipation of this current produces potentially 
observable hot spots in the star envelope. It also heats the planet 
and leads to a torque which drives the planet's orbit to evolve 
toward both circularization and a state of synchronization with 
the spin of the star. The net effect is the damping of the planet's 
orbital eccentricity. Around slowly (or rapidly) spinning stars, 
this process also causes rocky planets with periods less than a 
few days to undergo orbital decay (or expansion/stagnation) within
a few Myr. In principle, this effect can determine the retention
efficiency of short-period hot Earths. We also estimate the ohmic 
dissipation interior to these planets and show that it can lead to 
severe structure evolution and potential loss of volatile material 
in them.  However, these effects may be significantly weakened
by the reconnection of the induced field.
\end{abstract}

\keywords{magnetohydrodynamics (MHD) - planetary systems - planets and satellites: dynamical 
evolution and stability - planets and satellites: formation - planet-star interactions - 
stars: magnetic field}

\section{Introduction}
An important milestone in planetary astronomy is the discovery of a
Jupiter-mass planet, 51 Peg b \citep{may95}. Its
extraordinary 4 day orbital period rekindled a theoretical expectation
that protoplanets may undergo orbital decay \citep{gold80, lin86} 
as a consequence of their tidal
interaction with their natal disks. Today, more than 500 planets have
been discovered around nearby stars.  Among them, there is a pile up
of $\sim$100 planets with periods ($P$) less than a week and mass
($M_p$) two orders of magnitude larger than that of the Earth
($M_\oplus$). Transit observations of some of these planets indicate
that they have a radius and density comparable to that of Jupiter and
Saturn and are commonly referred to as hot Jupiters. 

These hot Jupiters, including 51 Peg b, were formed in presumably 
at some preferred locations (beyond the snow line) of their natal 
disks \citep{ida08, ida10}. After acquiring sufficient masses to open 
gaps in their natal disk, they undergo type II migration along 
with the viscous diffusion of their surrounding gas.  In order to 
account for their survival, we proposed that the migration of 
51 Peg b and other short-period gas giant planets may have 
stalled when they entered into the magnetospheric cavity of their 
host star during their infancy \citep{lin96}. Since 
protostellar disks are expected to be truncated within
the magnetosphere of their central stars (see below), once any 
protoplanet enters into this region, its migration would slow to 
a halt as the intensity of its tidal interaction with its 
nascent disk weakens.

The existence of magnetosphere around T Tauri stars was proposed
\citep{kon91} to account for the observed period distribution which
peaks around 8 days \citep{bou93}. If efficient angular
momentum flow between protostellar disks and the magnetosphere of
their central stars can enforce corotation at their interface where
the magnetic and viscous torques are balanced \citep{shu94},
we would infer kilogauss fields.
Today, Zeeman splitting of emission lines have been directly
measured for many T Tauri stars and these observations confirm the
presence of kilogauss fields on their surface \citep{joh07}. 
If this field strength represents that of a
dipole stellar field, the magnetospheric radius for T Tauri stars with
accretion rates in the range of $10^{-8} to 10^{-7} 
M_\odot$ yr$^{-1}$ would extend to regions beyond the
orbits of their close-in planets. The radius of this magnetospheric
cavity expands during the depletion of the disk gas and the decline of
the accretion flux through the disk.

Although the magnetospheric-cavity scenario provides a useful
qualitative model for the abundant population of short-period planets,
a detailed reconstruction of the observed period distribution requires
a quantitative determination of short-period planets' retention
efficiency. After they have entered the magnetospheric cavity of their
host stars or after they have been engulfed by the magnetospheric
cavity, protoplanets continue to interact with the stellar magnetic
field. It is not clear whether this process can induce planets to
undergo further orbital interaction.

In order to understand the nature of this physical mechanism, we carry
out a series of investigations.  In Paper I \citep{lai08},
we provided a general description of the relevant physical effects.
We first consider the retention of short-period gas giant planets
inside the stellar magnetosphere.  We decompose the stellar magnetic
field into a steady and a periodically modulating component and the
planet into a day and night side.  In this previous investigation, we
considered the periodically modulating component of the field which
can be due to either the planet's eccentric orbit or the star's
non-synchronous (with respect to the planet's orbital angular
frequency) and non-aligned (with respect to the star's magnetic poles)
spin.  On the night side of the planet where the magnetic diffusivity
is relatively high, this time-dependent field can permeate into the
planet's envelope and induce an AC current.  The ohmic dissipation of
this current not only heats the planet but also provides a torque
which drives the planet's orbit toward a state of circularization and
synchronization with the star's spin. We also found that close 
to the host star where the stellar magnetosphere is intense, 
ohmic dissipation can cause a planet's interior to heat up such 
that it expands and overflows its Roche lobe.
The gas flow from the planet to its host star via the inner Lagrangian
point also transfers angular momentum to the orbit of the planet 
\citep{gu03}.  This process can halt the migration of the planet
despite its loss of angular momentum as a consequence of its tidal
interaction with the disk and the host star and the direct torque
applied by the stellar magnetosphere on it.  The results of this
analysis will be presented elsewhere.

Another class of planets have been discovered with $M_p
\sim$ a few $M_\oplus$ and $P$ in the range of a few days to 2
months. In contrast to their Jupiter-mass siblings, these planets
probably have rocky or icy internal structures and are commonly
referred to as super-Earths \citep{may08, howard10}.  Recently 
Kepler mission led to the discovery of over 1200 planetary candidates.
Since they all have radii more than twice that of the Earth, they
may also be super-Earths, albeit their masses are yet to be determined.
These super-Earths too are probably formed at locations ranging from 
a fraction to several AUs from their host stars \citep{ida08, ida10}.
In contrast to the gas giants, super-Earths may
not have adequate mass to open a gap and undergo type II migration.
Nevertheless, they do tidally interact with their natal disk and undergo
type I migration, due to an imbalance between the torque they exert
on the disk at the Lindblad and corotation resonances both interior
and exterior to their orbits \citep{tan02}.

In most regions of the disk, type I migration is directed inward.
However, due to the corotation torque, there are migration barriers where
the orbital decay of isolated super-Earths may be halted 
\citep{mas06, par10}. These barriers occur at gas surface density
($\Sigma_g$) maximum near the snow line ($a_{\rm ice}$ where water
vapor condensate), at the inner edge of a dead zone ($a_{\rm dead}$ where 
the ionization fraction near the mid plane is too small to maintain
coupling between turbulent magnetic field and disk gas), and just
outside the magnetospheric cavity, $a_{\rm mag}$ \citep{kret07,
kret09}.

Since $\Sigma_g$ and the mid-plane temperature of the disk decline with
time, the location of these barriers also evolves with them.  During
the advanced stages of disk evolution, $\Sigma_g$ may be sufficiently
small that the dead zone essentially vanishes and the stalled embryos
near $a_{\rm ice}$ and $a_{\rm dead}$ resume their migration.
Assuming the field strength decreases slowly or remains unchanged, the
location of $a_{\rm mag}$ also expands beyond the orbital semimajor
axis of the super-Earths, $a_{\rm SE}$, during the transition from
classical to weak-line T Tauri phases on a timescale of $\sim 10$
Myr \citep{kret10}.  After the disk depletion, the stellar magnetic 
field and spin rate decrease \citep{skum72} on a timescale 
$\sim 100$ Myr \citep{sod93}.  During these evolutionary stages, the
location ($a_{\rm corote}$) where the frequency of the Keplerian
motion matches that of the stellar spin also evolve relative to
$a_{\rm SE}$.  Thus, we anticipate that some super-Earths are located
interior to the corotation radius while others are located beyond it.

In this paper, we consider the interaction between close-in super-Earths 
with the steady component of their host stars' magnetic field. Beside 
differences between their masses and radii, rocky planets have much
larger electric conductivity on their surface than the envelope and
atmosphere on the night side of close-in young gas giant planets.  The
dayside of gas giant planets is exposed to the stellar ionizing
photons and may have much higher ionization fraction and electrical
conductivities than their night side.  We will consider this more
complex aspect of the hot Jupiter problem elsewhere.  

The super-Earth problem we are considering here is analogous to the 
\textit{Echo} satellite (in the form of a conductor) moving relative to 
background (Earth) field line which was analyzed by Drell et al. (1965).
In general, an electric field is induced 
across the conductor (in directions orthogonal to the field and motion).  
However, the electric field must vanish (in its frame) inside a perfect 
conductor (with an infinite conductivity). On its surface, the electric
field generate a current which leads to an induced magnetic field. The 
induced field cancels the unperturbed field inside the conductor so that
there is no relative motion between the perfect conductor and the net
(unperturbed plus induced) field inside it.  Outside the moving conductor, 
the net field appears to wrap around it. At large distances from the
moving conductor, the induced field propagates away from it with the 
Alfven speed.

In a slightly different context, Goldreich and Lynden-Bell (1969) analyzed
the electrodynamic interaction between Io and Jupiter. They treated Io as 
a conductor moving in Jupiter's magnetosphere.  Because Io's conductivity 
is larger than that of Jupiter, it drags a flux tube of field lines. They 
showed that an electromagnetic field is induced across Io's surface.  
Provided the conductivity is high along and low across the field lines, 
the electric potential drop across them would propagate and be maintained 
along the field lines connecting Io and Jupiter with an Alfven speed. 
At the foot of the flux tube where it enters into Jupiter's atmosphere 
and envelope, conductivity across the field lines 
increases with the density of the surrounding gas such that the potential 
drop drives a DC current across the potential drop. Provided that the induced 
field can propagate back to Io before the unperturbed field slips through 
it, the current forms a close circuit. In this scenario, Io acts as a 
unipolar inductor.

In this paper, we apply the Goldreich and Lynden-Bell's model to the 
study of a super-Earth moving in its host star's magnetosphere. For 
computational simplicity, we neglect the planet's intrinsic magnetic field,
as we have done in the previous paper.  In the problem we are considering, 
a steady component of the stellar magnetic field is present 
regardless of the differential motion between the star's spin and the 
planet's orbit. In an asynchronous system, a flux tube of magnetic field 
between the planet and the star cannot be infinitely anchored on both 
entities.  If the planet's conductivity is 
smaller than that on the surface of the host star, the flux tube would 
tend to be anchored on the surface of the star along with all other 
unperturbed field lines and it would slip through the planet. 

In Section 2, we introduce qualitative discussion and schematic illustration 
to show in the limit that the electric conductivity in the super-Earth planet 
is higher (but not by an infinite amount) than that of its host star's 
envelope, the relative motion between the planet and the stellar magnetic 
field leads to an induced emf and a potential drop across the planet. 
Outside the planet, a flux tube of (unperturbed 
plus induced) magnetic fields would appear to be approximately anchored 
on the planet. In the tenuous regions between the planet and its host star, 
the conductivity along the field lines is much higher than that across them 
and the electric current flows freely to maintain constant electric potential 
along them. In the absence of field reconnection, Alfven waves propagate to
infinity on open lines and the electric current flows to the surface of the
host stars along closed field lines.  Due to the finite resistance of the 
surrounding (stellar) gas, the foot of the flux tube on the surface of the 
host star slips through the stellar atmosphere and
the electrical potential drop across the foot of the flux tube drives 
an electric current with an associated rate of ohmic dissipation.  

Based on the above qualitative scenario, we construct a quantitative model
in this paper.  In Section 3, we introduce the values of the different parameters 
we use in the numerical applications and derive the analytical expressions 
for the induced electric field, intensity, ohmic dissipation, and torque. 
In our numerical applications, we adopt a set of fiducial physical parameters 
for a rocky super-Earth planet with two Earth radii, $2 R_\oplus\approx 
1.4 \times 10^{7}$m on a 3 day circular orbit ($a \approx 6 \times 10^{9}$m 
around a T Tauri star with a mass equals to that of the Sun ($1M_\odot \approx 2 \times 
10^{30}$kg) and a radius twice that of the Sun $R_\ast= 2 R_\odot 
\approx 1.4 \times 10^{9}$m.  We also assume that it has a solar luminosity 
($1 L_\odot$), a surface temperature of $T_\ast= 4 \times 10^3$K, and a dipole 
field with a strength 0.2 T (i.e. $ 2 \times 10^3$ G) on the stellar surface 
(which corresponds to a magnetic dipole strength of $5.4 \times 10^{33} 
{\rm Am}^{2}$).  

With these parameters, we first analyze two limiting cases of rapid and 
slow stellar spin.  We discuss the condition of validity of the model in 
Section 4 and derive in Section 5 the expressions and values of the resistances and 
Alfven speed. We will consider more general sets of model parameters 
elsewhere. In Section 6, we calculate the values of the induced intensity, 
ohmic dissipation, and torque, and discuss the relevance of these values. 
In the context of planetary migration in the presence of their natal disk, 
we show that if the planet orbits around the host 
star outside its corotation radius (i.e. the Keplerian frequency 
of the planet's orbit is smaller than the stellar spin frequency), the 
net torque associated with this induced current would provide an adequate 
rate of angular momentum transfer to balance against the rate of tidally 
induced angular momentum loss by the rocky planet to the disk.  In the
limit that the planet is inside its host star's corotation radius, the
planet's orbit would continue to decay.  Finally, we summarize our results
and discuss their implications in Section 7.

\section{Qualitative Illustration and Description of the Phenomena and 
Brief Outline of the Calculation}

We consider a rocky planet with a mass of several $M_\oplus$ moving 
in the dipole magnetic field of the star it is orbiting. The relative 
motion of such a conductor in an external stellar magnetic field 
generates an induced emf across the planet. There are two complementary effects 
(see paper 1) described by the complete magnetohydrodynamic (MHD) equation: the 
diffusion of the magnetic field in the planet and the magnetic induction 
(with its associated drag).

The MHD equation describing the electrodynamics
of the planet in the stellar field can be written as
\begin{equation}
\frac{\partial \textbf{B}}{\partial t} 
= \nabla \wedge \left(\upsilon \wedge \textbf{B}\right)  
- \nabla \wedge \left(\eta \nabla \wedge\textbf{B} \right)
\label{MHD}
\end{equation}
where $\eta$ is the magnetic diffusivity (which is equal to $(\mu_{0}\sigma)^{-1}$ where $\sigma$ is the electric conductivity).

In Paper 1, we focused on the diffusion 
of the stellar magnetic field inside a hot Jupiter. In order to do so, 
we considered the periodic component of the stellar magnetic field felt 
by a planet when the axis of the stellar magnetic moment is not aligned 
with the planetary orbital axis, or when the planet is on an eccentric 
orbit. The drag of the magnetic field by the planet due to induction 
can be neglected if the planet's orbit co-rotates with the star's
spin or if the electrical conductivity of the planet is low compared 
with that of the outer layers of the star (which is the case at
least in the night side of a hot-Jupiter).  

In this limit, the diffusion of the stellar magnetic field inside 
the planet is modulated by the electric conductivity (inversely 
proportional to the magnetic diffusivity) profile in the planet. 
A somewhat higher electric conductivity in the planet would tend 
to decrease the penetration depth of the stellar magnetic field and 
the volume where the electric current induced by the field can be 
dissipated, but it would also increase the volumic ohmic (power)
dissipation rate. Likewise, a lower conductivity in the planet 
would enable the stellar magnetic field to penetrate deeper into it, 
albeit the induced current also encounters a lower volumic ohmic 
dissipation rate. Consequently, we found that, the total ohmic 
dissipation rate over the entire planet does not change significantly 
over a reasonable range of electric conductivity for a hot-Jupiter 
(neglecting the effect of photo-ionization in its atmosphere).
 
In the present paper, we describe the induction (and associated 
"drag" of the field lines) which was neglected in paper I.  For 
simplicity, we neglect the damping of the magnetic field in the 
planet associated with the diffusion term and 
focus on the case where the planet is able to significantly drag 
the stellar magnetic field lines which are enclosed in the flux 
tube that passes through the planet. 

Throughout this paper, this "flux tube which passes through 
the planet" is simply referred to as "the flux tube" (we are 
interested in the part that extends between the interior 
of the planet and above the surface of the star). The ``foot 
of the flux tube" refers to that part of the flux tube which 
extends below the surface of the star for a distance $D_{pn}$ 
(the subscript "pn" refers to penetration) to be estimated below. 

Between the planet 
and the surface of the star, the volumic current flows along 
the flux tube (parallel to the magnetic field lines--the 
electrons in fact gyrate around the magnetic field lines), 
but the volumic current crosses the flux tube in the planet 
and at the foot of the flux tube (perpendicular to 
the magnetic field lines in the stellar atmosphere). 
Figures 1 and 2 present the general overview of the system.

\begin{figure}
\includegraphics[scale=0.7]{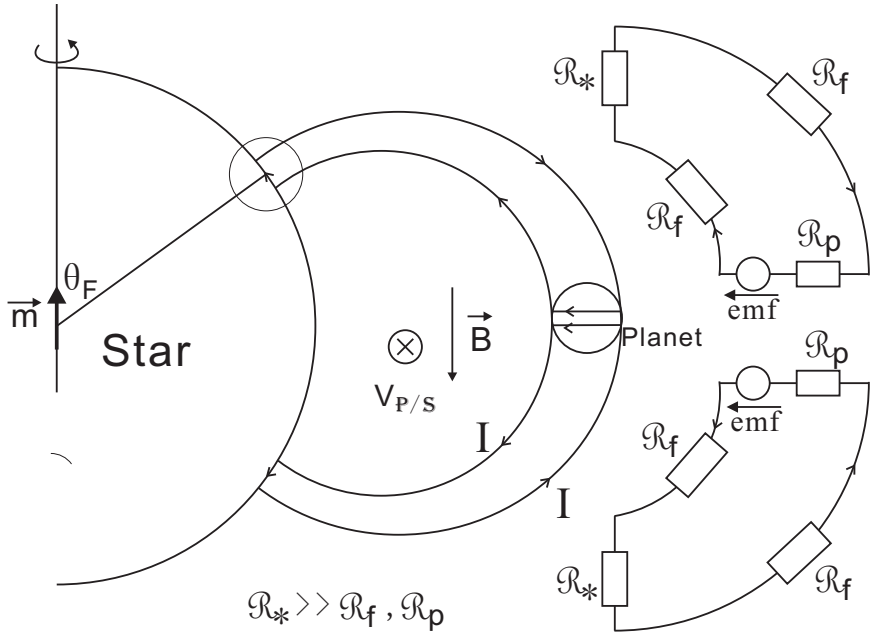}
\label{overview}
\caption{Schematic illustration of the planet-magnetosphere 
interaction system. The circuit diagram idealizes the basic 
physics which is described in the text.  In this context,
a planet with a non-negligible motion (into the plane of the 
diagram) relative to the stellar magnetosphere induces an emf.
At the location of the planet, the direction of the unperturbed
stellar dipole field is pointing downward.  The potential 
difference across the flux tube generate a current with a 
flux which is primarily determined by the electrical resistivity 
in the atmosphere of the host star. Arrows indicate the flow 
direction of the current.}
\end{figure}

\begin{figure}
\includegraphics[scale=0.7]{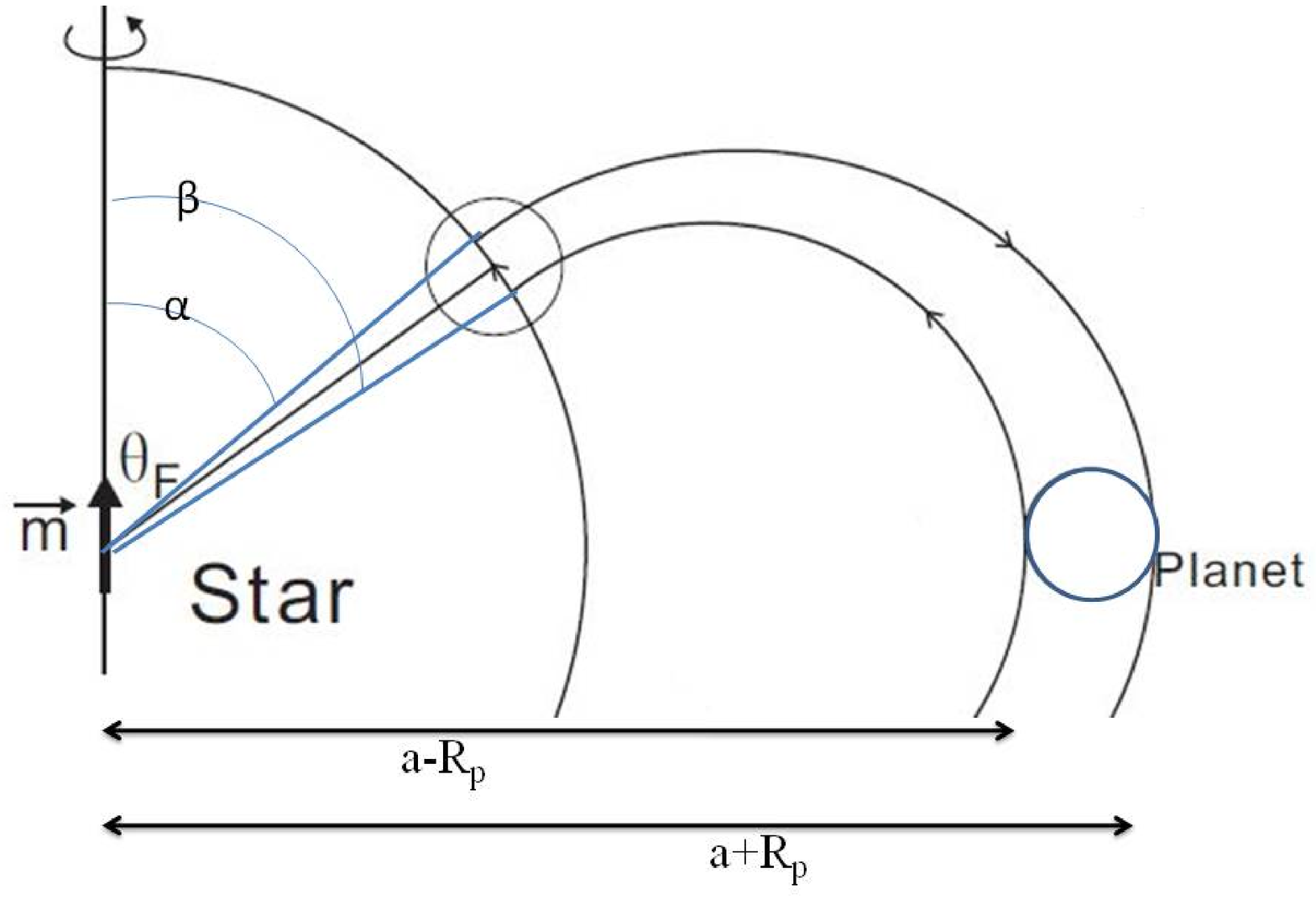}
\caption{Main parameters in the calculation of $y_{1}$ and $y_{2}$ which defines 
the geometry of the foot of the flux tube}. 
\label{geometry}
\end{figure}

In this model, we make a distinction between the regular 
(unperturbed) stellar dipole magnetic field (magnetosphere) which co-rotates 
with the stellar spin 
and the field lines which define the flux tube (composed of 
both the stellar field in the planet and the 
induced flux tube) which appear
to be dragged along with the planet and thus move relative to 
the rest of the stellar magnetosphere. This drag is significant 
when the electric conductivity of the planet is large compared to 
that of the outer layers of the star \citep{pid68}.

\begin{figure}
\includegraphics[scale=0.8]{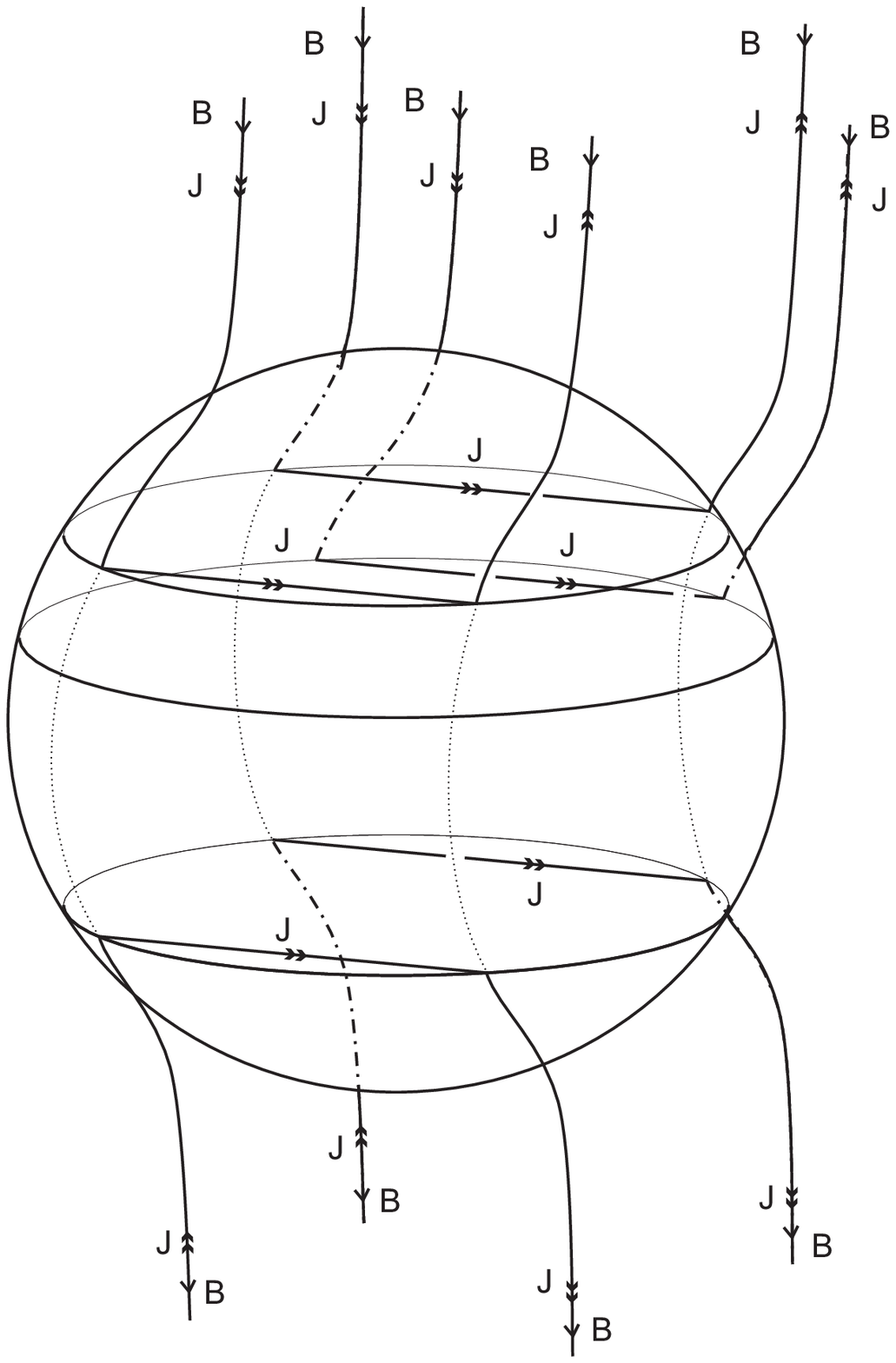}
\caption{External and induced field around a planet moving relative
to the magnetosphere of its host star. The planet is moving out of the
plane and toward the lower left side of the illustration.  Arrows in 
this idealized cartoon 
illustrate the direction of current flow across the potential drop due 
to the induced emf. The current also lead to an induced field which 
distorts the field near the planet.  Information on the induced
emf propagates along the field lines in the direction of the host star.}  
\label{fig:planet}
\end{figure}

A large conductivity in the planet's interior would 
lead to a large surface current which induces a field and cancels 
the external (unperturbed stellar dipole) field. With small magnetic 
diffusivity, the planet's interior would appear to be shielded from 
any time dependent external magnetic field. We, therefore, consider 
only the time-independent component of the stellar magnetic field.
In this model, the time-independent component of the stellar magnetic
field permeates the planet. The motion of the planet relative to the 
stellar magnetosphere induces an electric field $\mathcal{E}$, an induced 
volumic electric current $\mathcal{J}$, and an induced difference 
of potential U across the planet's diameter. 
In Figure \ref{fig:planet}, we provide a schematic illustration 
on the field lines and current inside the planet. 

For the flux tube between the planet and its host star, we adopt the
assumption of high electric conductivity along the magnetic field lines 
and low conductivity across them which was introduced by Goldreich \&
LyndenBell (1969) for the system Io-Jupiter. In this case, the difference 
of electric potential (induced by the planet's relative motion with respect
to the magnetic field of its host star) drives an electrical current out 
of the planet, along the flux tube, across its foot on the atmosphere 
of the star, and back to the planet along its other half (Figure 1). 

In the limit of negligible electrical conductivity in the direction normal 
to the flux tube, the electric current can only cross the field lines 
in the planet or in the atmosphere of the host star. The assumption of high 
electric conductivity along the magnetic field lines also implies 1) that 
the difference of potential U across the planet's diameter is transmitted 
along the flux tube without significant drop in potential and 2) that the 
plasma enclosed in the flux tube is dragged along by the motion of the 
magnetic field lines. An electric circuit is therefore created, where
the flux tube acts as electric wires, the planet as a unipolar inductor 
with internal resistance $\mathcal{R}_{p}$, and the foot of the flux tube 
as the largest resistance. 

We see that there are in fact two circuits (see Figure 1). The first one 
is composed of the foot of the flux tube on the northern 
hemisphere of the star, its corresponding flux tube, and the northern 
hemisphere of the planet. The second is 
equivalent and symmetric to the first one (the plane of symmetry being 
the plane of the planetary disk). Except when explicitly stated, the 
calculations (current, resistances, ohmic dissipation, torques, etc.) 
describe only one circuit (the northern one). 

In an electric circuit composed of a generator (with an emf $\int \mathcal{E} {\rm d}l $ 
and resistance $\mathcal{R}_{g}$ over a length scale $l$) and other 
resistances along the circuit $\mathcal{R}$ (here primarily the resistance 
of the foot of the flux tube), the intensity of the current $I$ is 
determined by $\int \mathcal{E} {\rm d} l - \mathcal{R}_{g} I \approx 
\int \mathcal{E} {\rm d} l = U = \mathcal{R} I$. With the 
parameters we have adopted here, we show in Section 5 that the resistance 
along the flux tube $\mathcal{R}_{\rm tube}$ and the resistance across 
the planet $\mathcal{R}_{\rm p}$ are small compared to that 
across the foot of the flux tube on the star $\mathcal{R}_\ast$. In
this limit, 1) the potential drop across the planet with a radius $R_p$ 
is $U \sim 2 \mathcal{E} R_p$, 2) the magnitude $U$ is approximately 
constant along each field lines in the flux tube between the planet 
and its host star because the resistance of the tube $\mathcal{R}_{\rm tube}$ 
and the induction are negligible, and 3) the total current is determined 
by the largest resistance along the circuit, {\it i.e.} that at the foot 
of the flux tube in the stellar atmosphere.

\section{Description of the Model: Values of the Parameters, Geometry, 
Analytical Expressions, and Equation of Evolution of the Planet's Orbit}

\subsection{Values of the Parameters for a Fiducial Model}
Except when explicitly stated otherwise, we consider a system composed of a 
super-Earth orbiting closely a young T-Tauri star with a time-independent 
magnetic dipole. We assume that the magnetosphere co-rotates with the star, 
and adopt the following numerical values (SI units) for the parameters intervening in 
the model: 

For the star we adopt the following model parameters: \\
Temperature of the isothermal outer layer: $T_\ast=4000$ K, \\
Radius: $R_{\star} \approx 2 R_{\odot} \approx 1.4\times 10^{9}$ m, \\ 
Mass: $M_{\star} \simeq M_{\odot} \simeq 2 \times 10^{30}$ kg,  \\
Opacity at the photosphere: $\kappa \simeq 3$ ${\rm m}^{2}{\rm kg}^{-1}$ 
(which is equivalent to taking a surface pressure of about 15 Pa), \\
Magnetic dipole strength: $m=5.4 \times 10^{33}$ ${\rm Am}^{2}$, which 
corresponds to a magnetic field of 0.2 T (Tesla) $\equiv 2
\times 10^3$ Gs (Gauss) at the stellar surface \citep{yang08}, and \\
Spin period: 8 days (slow-rotator) or 0.8 days (fast-rotator). \\
We will consider more general stellar models elsewhere. 

For a super-Earth, we consider the following case: \\
Radius: $R_{p} \simeq 2R_{\oplus} \simeq 1.4 \times 10^{7}$ m, and \\
Semi-major axis: $a \simeq 0.04$ AU $\simeq 6 \times 10^{9}$ m 
(which corresponds to a period of 3 days).\\
In the electrodynamics of super-Earths, the magnitude of $M_p$ 
does not enter explicitly (it does implicitly through the radius) the calculation of the torque, 
albeit their orbital evolution timescale  
does depend on it (for example, see Equation (\ref{EvoPlanet}).  
The linear speed in a frame co-rotating with the star 
($\upsilon_{p/s} = (\omega_{p}-\omega_{\ast})a$) of such 
a planet orbiting a star rotating slowly is $9 \times 10^{4} {\rm m s}^{-1}$ 
and $4 \times 10^{5} {\rm m s}^{-1}$ around a star rotating fast (these are 
the absolute values).

\subsection{Length and Width of the Foot of the Flux Tube in a 
Spherical Approximation}

As defined above, the "flux tube" refers to the flux tube composed of the 
field lines of the magnetosphere that pass through and are dragged 
along by the planet. This flux tube connects the planet and the surface 
of the star. and its foot penetrates into the star to a depth which will 
be determined later. 

The stellar magnetic field has the geometrical structure of a magnetic 
dipole field. Thus, the foot of the flux tube at the stellar envelope 
(the circled area in figure \ref{geometry}) is, at the first order 
in $\sqrt{{R_{\ast}}/{a}}$, an ellipse which axes have respective lengths 
\begin{eqnarray}
\label{y1 and y2}
y_{1} &=& \left(\frac{R_{p}}{s}\right) 
\left(\frac{R_{\ast}}{a}\right)^{3/2}, \\
y_{2} &=& 2 R_{p} \left(\frac{R_{\ast}}{a}\right)^{3/2}, \\
sin \theta_{F} &=& \sqrt{\frac{R_{\ast}}{a}} \\
cos \theta_{F} &=& \sqrt{1-\frac{R_{\ast}}{a}} \equiv s, 
\end{eqnarray}
where $\theta_{F}$ is the angle between the stellar spin axis and the 
location of the foot of the flux tube. When the current $\textbf{J}$
crosses the foot of the flux tube in the stellar envelope, it covers a 
length $y_{1}$. In the rest of the paper, we take $cos \theta_{F}$ is roughly equal to 
1. We represent it with the symbol s in analytical equations and take it to be equal to 1
in numerical applications. 

\begin{figure}
\includegraphics[scale=0.8]{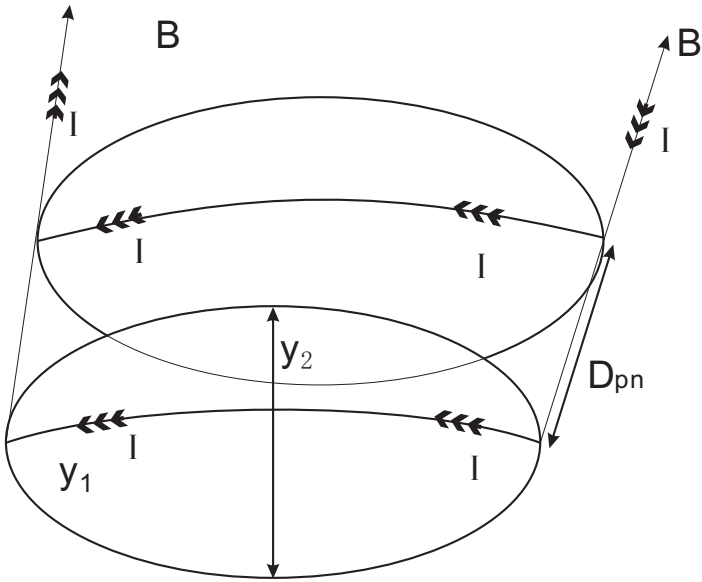}
\caption{Penetration of the flux tube in the stellar atmosphere.
Potential difference across the planet would be maintained at its
foot print on the surface of its host star if there is sufficient
time for the Alfven waves to transit this information.  Arrows 
indicate the flow direction of the electrical current.  The values 
of $y_1$ and $y_2$ are given by Equations (\ref{y1 and y2}). The
top circle represents the stellar surface and
the penetration depth $D_{pn} = R_\ast-r_{pn}$ is evaluated
in Section \ref{sec:sigmastar}.}
\label{fig:foot}
\end{figure}

In Figure \ref{fig:foot}, we zoom in on the foot of the flux tube at the 
stellar atmosphere. In order to derive $y_{1}$ and $y_{2}$ (at the first 
order in ${R_{\ast}}/{a}$), we first solve $\mathcal{B} \wedge dl = 0$ 
and obtain $sin \alpha = \sqrt{\frac{R_{\ast}}{a+R_{p}}}$ and $sin \beta 
= \sqrt{\frac{R_{\ast}}{a-R_{p}}}$ with $\alpha$ and $\beta$ defined in 
Figure \ref{geometry}. We then write $y_{1}=R_{\ast}(\alpha -\beta)$ 
and $y_{2}=\frac{2R_{p}}{2\pi a} 2 \pi R_{\ast} sin \theta_{F}$.

For a super-Earth (using the model parameters listed above), we find
$y_{1}=1.6 \times 10^{6}$ m and $y_{2}=3.2 \times 10^{6}$ m. For a 
hot-Jupiter, we would typically need to multiply these values by a 
factor $10$. The height of the foot of the flux tube is derived below in 
Section \ref{sec:depth}. The numerical applications are for $s=1$. 
For semi-major axes comparable to the stellar 
radius, the multiplicative factor $(1-{R_{\ast}}/{a})^{-1/2}$ 
in $y_{1}$ would significantly affect the length of the foot of 
the flux tube.

\subsection{Induced Difference of Potential}
The planet is a conductor moving in the stellar magnetosphere with 
relative linear speed $\upsilon_{p/s}$. Modeling the stellar magnetic 
field as the one created by a magnetic dipole (of magnetic moment m), 
the magnitude of the induced electric field in the planet $\mathcal{E}_{p}$ is 
\begin{equation}
\mathcal{E}_{p}=\upsilon_{p/s} \mathcal{B}_{\ast}(a) = 
(\omega_{p}-\omega_{\ast})a \frac{\mu_{0}m}{4\pi a^{3}}
\label{eq:vcrossb}
\end{equation}
with $\mathcal{B}_{\ast}(a)$ represents the stellar magnetic field at 
the location of the planet. Numerical applications for a super Earth 
give $\mathcal{E}_{p}=240$ V ${\rm m}^{-1}$ (slow-rotator), $1000$ 
V${\rm m}^{-1}$ 
(fast rotator) (a slow or fast rotator depends on the spin frequency 
of the star, as described in section 3.1).

The magnitude of the difference of potential U (or emf) generated across 
the planet is thus 
\begin{equation}
U=2R_{p} \mathcal{E}_{p} = 2R_{p}(\omega_{p}-\omega_{\ast})a 
\frac{\mu_{0}m}{4\pi a^{3}}.
\end{equation}
For the super-Earth models under consideration, $U=6.7 \times 10^{9}V$ 
(slow rotator) and $2.8 \times 10^{10}V$ (fast-rotator).
This difference of potential is transmitted across the flux tube (with 
the assumption of infinite conductivity along the flux tube that passes 
through the planet) and generates a uniform electric field 
$\mathcal{E}_{\ast}$ in the stellar envelope (see Figure 1)
at the foot of the flux tube
\begin{equation}
\mathcal{E}_{\ast} = \frac{U}{y_{1}}=2 \frac{\mu_{0}m}
{4\pi a^{3}} (\omega_{p}-\omega_{\ast})a \left(\frac{a}{R_{s}}\right)^{3/2} s.
\end{equation}

The value of $\mathcal{E}_{\ast}$ does not depend on the radius of the 
planet, and for our values of the parameters for a young T Tauri star, 
$\mathcal{E}_{\ast}=4.2 \times 10^{3} $V$m^{-1}$ (slow rotator), $1.8 
\times 10^{4} $V$m^{-1}$ (fast rotator). 

\subsection{Analytical Expressions: Intensity in the Circuit, 
Ohmic Dissipation, and Torque in the Planet and Star}
\label{sec:torque}

The induced current I is given by
\begin{equation}
\label{I}
I = \int_z \int_y \mathcal{J} dydz = \mathcal{E}_{\ast} y_{2} 
\int_z \sigma_{\ast}(z) dz 
= U \frac{y_{2}}{y_{1}} \int_z \sigma_{\ast}(z) dz
=4 R_{p} (\omega_{p}-\omega_{\ast})a \frac{\mu_{0}m}{4\pi a^{3}}s 
\int_z \sigma_{\ast}(z) dz.
\end{equation}
In the previous equation, $\mathcal{J}$ is the volumic electric current 
in the stellar atmosphere at the foot of the flux tube (induced by U), 
y varies from 0 to $y_{2}$ (the width of the foot of the flux tube), 
and z varies from $r_{pn}$ (the radius to which the flux tube can 
penetrate into the stellar atmosphere) to $R_{\ast}$ and 
$\frac{y_{2}}{y_{1}}=2s$ (see Figure \ref{fig:foot}). In this circuit,
the total resistance is the sum of that across the planet, the foot print
of the flux tube on the stellar surface, and along the flux tube.  Here 
we consider only the largest contribution and neglect that across the planet.
In Section \ref{sec:sigmastar}, we determine the magnitude of $r_{pn}$ and evaluate 
\begin{equation}
\Sigma = \int_{r_{pn}} ^{R_{\ast}} \sigma_{p}(z) dz
\end{equation}
such that $I=2U\Sigma s$. In the above equation, $\sigma_{p}(z)$ is the 
local Pedersen conductivity (see Equation \ref{eq:sigmap}). The total 
resistance of the stellar atmosphere at the foot of the flux tube is 
\begin{equation}
\mathcal{R}_{\ast}=\frac{U}{I}= \frac{y_{1}}{y_{2}}\frac{1}{\Sigma} = 
\frac{1}{2\Sigma s}.
\label{Rstar}
\end{equation}

The total ohmic power dissipation in the stellar atmosphere at the foot of 
both flux tubes (one for each hemisphere, thus the multiplicative factor 2) $\mathcal{P}_{\ast}$ and in the planet 
$\mathcal{P}_{p}$ are 
\begin{eqnarray}
\mathcal{P}_{\ast} &=& 2 \mathcal{R}_{\ast} I^{2} = 2UI= 4 U^{2}\Sigma s= 
16 R_{p}^{2} (\omega_{p}-\omega_{\ast})^{2}a^{2} 
\left(\frac{\mu_{0}m}{4\pi a^{3}}\right)^{2} \Sigma s \\
\mathcal{P}_{p} &=& 2\mathcal{R}_{p} I^{2}= 2\mathcal{R}_{p} \left(2 U 
\Sigma s \right)^{2}= \mathcal{P}_{\ast} \frac{\mathcal{R}_{p}}
{\mathcal{R}_{\ast}}
\end{eqnarray}
where $\mathcal{R}_{\ast}$ is the resistance of the foot of the flux 
tube (on one hemisphere), and 
$\mathcal{R}_{p}$ is the resistance of one hemisphere of the planet. 
 Since the resistance in the planet is considerably smaller than that
in the star, most of the power is dissipated in near the foot of the 
flux tube on the surface of the star.  Note that the magnitude of both
$\mathcal{P}_{\ast} $ and $\mathcal{P}_{p}$ is determined by $I$ and 
$ \mathcal{R}_{\ast}$.

The total torque (for both circuits, one circuit for each hemisphere) 
due to the Lorentz force (the axis is the stellar spin axis) on the 
star (equal in absolute value to that on the planet) is
\begin{eqnarray}
\mathcal{T}_{\ast} &=& 
2 \int_{x} \int_{y} \int_{z} r \wedge 
(\mathcal{J} \wedge \mathcal{B}) dx dy dz 
=2(R_{\ast} sin \theta_{F}) (y_{1}I) (\frac{2 \mu_{0}m cos \theta_{F}}{4\pi R_{\ast}^{3}})
=4R_{p}a\frac{\mu_{0}m}{4\pi a^{3}}I \textbf{e}_{z}, \\
\mathcal{T}_{\ast} &=& 16 R_{p}^{2}a^{2}(\omega_{p}-\omega_{\ast}) 
\left(\frac{\mu_{0}m}{4\pi a^{3}}\right)^{2} 
\Sigma s \textbf{e}_{z}
\label{torque}
\end{eqnarray}
where $s=cos \theta_{F}$ as defined in (\ref{y1 and y2}) and I is the integral of the volumic current across a cross section (we take an averaged view of the volumic current rather than 
determining its complex geometry inside the planet). We have calculated here the total ohmic dissipation and torque (i.e. for both hemispheres). 

\subsection{Equation of Evolution of the Stellar Spin and Planet's Orbital 
Angular Velocity and Semimajor Axis}
The torque on the planet $\mathcal{T}_{p}$ is equal and opposite that 
on the star $\mathcal{T}_{\ast}$.  Consequently, the semimajor axis 
of a super-Earth on a circular orbit evolves at a rate
\begin{equation}
{\dot a} = \frac{2 a}{H_p} \mathcal{T}_{p}
\label{EvoPlanet} 
\end{equation} 
where the total angular momentum of the planet's orbit is 
$H_p = M_p a^2 \omega_p$. Since the total angular momentum of the
system is conserved, the changing rate of the stellar spin is
\begin{equation}
{\dot \omega_\ast} = \frac{\mathcal{T}_\ast} {c_\ast M_\ast R_\ast^2}
\label{EvoStar}
\end{equation}
where $c_\ast \simeq 2/5$ is the inertial constant of the star.  According to 
the above expression, the planet would undergo orbital decay and
its host star would spin up if it is inside corotation (or 
equivalently if $\omega_p > \omega_\ast$).  Similarly, the planet's
orbit would expand and its host star would spin down if it is outside
corotation.

The planet's orbital frequency $\omega_{p}$ is related to the semi-major 
axis $\omega_{p}=\frac{\sqrt{GM_{\ast}}}{a^{3/2}}$. Using the expressions 
calculated for the torques and Equations (\ref{EvoPlanet}) and 
(\ref{EvoStar}), we find
  
\begin{equation}
\stackrel{\centerdot}{\omega}_{\ast}=\frac{M_{p}(GM_{\ast})^{2/3}}
{3C_{\ast}M_{\ast}R_{\ast}^{2}} 
\frac{\stackrel{\centerdot}{\omega_{p}}}{\omega_{p}^{4/3}}
\approx 3 \times 10^{-10} \frac{\stackrel{\centerdot}{\omega}_{p}}
{\left(\omega_{p} \stackrel{\sim}{s}\right)^{4/3}}
\end{equation}
where the numerical application was for a planet with 10 Earth 
Masses ($M_{p}=6 \times 10^{25}$kg) and $\rm \stackrel{\sim}{s}$ represent second. 
Therefore, we can estimate the variation of $\omega_{\ast}$ during 
the evolution of the planet's migration
\begin{equation}
\mid \bigtriangleup \omega_{\ast} \mid = - \frac{M_{p}(GM_{\ast})^{2/3}}
{C_{\ast}M_{\ast}R_{\ast}^{2}} 
\mid \bigtriangleup \left( \frac{1}{\omega_{p}^{1/3}} \right) \mid
\end{equation}
which is negligibly small.

We can thus consider that the star's angular velocity is roughly 
unaffected by this transfer of angular momentum. Using Equation 
(\ref{EvoPlanet}), $a^{3}=\frac{GM_{\ast}}{\omega^{2}_{p}}$, 
$\frac{\stackrel{\centerdot}{a}}{a}=\frac{-2}{3}\frac{\stackrel{\centerdot}{\omega_{p}}}{\omega_{p}}$, and the expression for the torque on the planet 
(equal to $-\mathcal{T}_{\ast}$ with $\mathcal{T}_{\ast}$ 
given in Equation (\ref{torque})) we find
\begin{equation}
\frac{\stackrel{\centerdot}{\omega_{p}}}{\omega_{p}^{4}(\omega_{p}
-\omega_{\ast})} = 48 \frac{R_{p}^{2}}{M_{p}} \left(\frac{\mu_{0}m}{4\pi}\right)^{2} 
\frac{\Sigma s}{(GM_{\ast})^{2}} = \alpha
\end{equation}
where $\omega_{\ast}$ is constant and $\alpha=48 \frac{R_{p}^{2}}{M_{p}} \left(\frac{\mu_{0}m}{4\pi}\right)^{2} 
\frac{\Sigma s}{(GM_{\ast})^{2}}$ (of unit $s^{3}$), and s in the numerator of the expression for $\alpha$ is defined as in \ref{y1 and y2}. 
The previous equation becomes 
\begin{equation}
-\frac{1}{\omega_{\ast}}\frac{\stackrel{\centerdot}{\omega_{p}}}{\omega_{p}^{4}}
-\frac{1}{\omega_{\ast}^{2}}\frac{\stackrel{\centerdot}{\omega_{p}}}{\omega_{p}^{3}}
-\frac{1}{\omega_{\ast}^{3}}\frac{\stackrel{\centerdot}{\omega_{p}}}{\omega_{p}^{2}}
-\frac{1}{\omega_{\ast}^{4}}\frac{\stackrel{\centerdot}{\omega_{p}}}{\omega_{p}}
+\frac{1}{\omega_{\ast}^{4}}\frac{\stackrel{\centerdot}{\omega_{p}}}{(\omega_{p}-\omega_{\ast})}
= \alpha .
\end{equation}

After integration, we find

\begin{equation}
\frac{\Omega^{3}(t)}{3} + \frac{\Omega^{2}(t)}{2} + \Omega(t) + 
ln \left( \mid 1-\Omega(t) \mid \right) = 
\frac{\Omega^{3}_{i}}{3} + \frac{\Omega^{2}_{i}}{2} + \Omega_{i} + 
ln\left(\mid 1-\Omega_{i} \mid \right) + \alpha \omega_{\ast}^{4}(t-t_{0}) 
\label{omegaPt}
\end{equation}
where we define $\Omega(t) \equiv {\omega_{\ast}}/{\omega_{p}(t)}$ 
and $\Omega_{i} \equiv \Omega(t=t_{0})$.

Near co-rotation (i.e. $\Omega \simeq 1$), the ln function is dominant 
and we find
\begin{equation}
ln \left( \mid 1-\Omega(t) \mid \right) = ln\left(\mid 1-\Omega_{i} 
\mid \right) + \alpha \omega_{\ast}^{4} (t-t_{0}).
\label{EvoOmegaPlanet}
\end{equation}
If $\omega_{p}$ is greater than $\omega_{\ast}$, then
\begin{equation}
\frac{\omega_{p}(t)}{\omega_{\ast}} = \frac{1}{\Omega(t)} 
= \left[1 - \left(1-\frac{\omega_{\ast}}{\omega_p(t_{0})}\right)
exp\left(\alpha \omega_{\ast}^{4} (t-t_{0})\right)\right]^{-1}.
\end{equation}

Similarly, if $\omega_{p}$ is less than $\omega_{\ast}$, then
\begin{equation}
\frac{\omega_{p}(t)}{\omega_{\ast}} = \frac{1}{\Omega(t)} 
= \left[1 + \left(\frac{\omega_{\ast}}{\omega_p(t_{0})} - 1 \right)
exp\left(\alpha \omega_{\ast}^{4} (t-t_{0})\right)\right]^{-1}.
\end{equation}

and the planet-star system approaches co-rotation with a timescale 
\begin{equation}
\tau= \frac{1}{\alpha \omega_{\ast}^{4}} =\frac{M_{p}(GM_{\ast})^{2}}{48\omega_{\ast}^{4}R_{p}^{2}\Sigma s}
\left(\frac{\mu_{0}m}{4\pi}\right)^{-2}. 
\end{equation}
In the general case, $\omega_{p}(t)$ follows Equation (\ref{omegaPt}), 
or written differently:
\begin{equation}
\omega_{p}^{3}(t)\left[f(t) - ln \mid 1 - \frac{\omega_{\ast}}{\omega_{p}(t)} 
\mid \right]
-\omega_{\ast} \omega_{p}^{2}(t) - \frac{\omega_{\ast}^{2}}{2}\omega_{p}(t)
=\frac{\omega_{\ast}^{3}}{3}
\end{equation}
where $f(t)=A + B(t-t_{0})$ with $A={\Omega^{3}_{i}}/{3} + 
{\Omega^{2}_{i}}/{2} + \Omega_{i} + ln\left(\mid 1-\Omega_{i} 
\mid \right)$ and $B={\alpha \omega_{\ast}^{4}}.$

In order to get an equation of evolution of the semi-major axis, one 
can replace in Equation (\ref{EvoOmegaPlanet}, near co-rotation) or 
(\ref{omegaPt}, in the general case) $\Omega_{p}(t)$ by $\left({a(t)}/
{a_{c}}\right)^{3/2}$ and 
$\Omega_{i}$ by $\left({a(t_{0})}/{a_{c}}\right)^{3/2}$ with $a_{c}$ 
the co-rotation radius. Near co-rotation, we find
\begin{equation}
a (t) = a_c \left[ 1 - \left[ 1 - \left( a(t_0) \over a_c \right)^{3/2} 
\right] {\rm exp} \left( { t - t_0 \over \tau} \right) \right]^{2/3}
\end{equation}
for the case where a is smaller than $a_{c}$.

\section{Condition for the validity of the Model: $t_{A} \leq t_{max}$}
\label{sec:validity}
\begin{figure}
\includegraphics[scale=0.5]{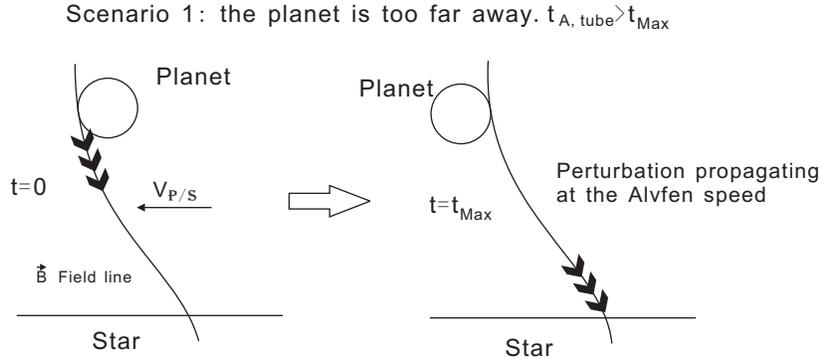}
\caption{Propagation of the Alfven wave between a relatively distant
planet and its host star.  The planet's motion relative to the stellar 
field induces a potential drop across the flux tube in the proximity 
of the planet.  This information propagates along the flux tube toward 
the host star with an Alfven speed.  Due to finite diffusion and the 
relative motion between the planet and the stellar magnetosphere,
the net field lines also slip through the planet.  In this illustration
the timescale required for the Alfven wave to reach the host star
is long compared with that required for the slippage of the field.
The circuit is not established in this case.}
\label{fig:move1}
\end{figure}

\begin{figure}
\includegraphics[scale=0.5]{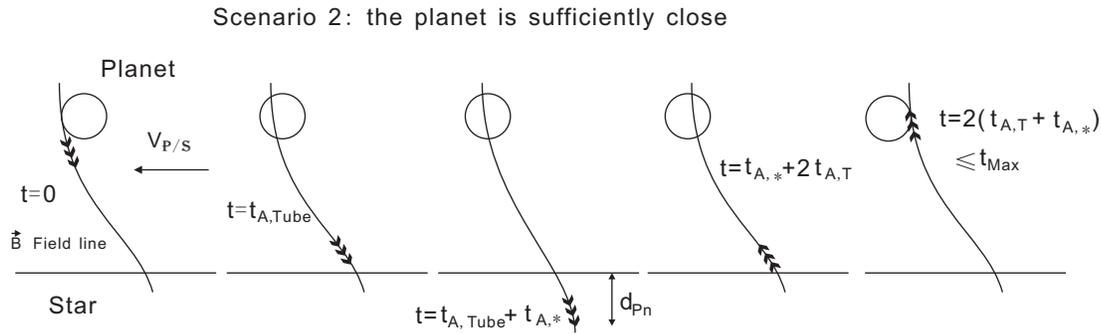}
\caption{Necessary condition for a complete unipolar inductor 
circuit.  Similar to the illustration in Figure \ref{fig:move1}, the
potential drop across the planet propagates along the flux tube toward
the planet with an Alfven speed.  In this illustration, the planet is
sufficiently close to its host star that the potential drop can be
established on the surface of the host star before the fields slip 
through the planet.  This potential drop induces a current which is
determined by the resistivity on the stellar surface.  In this case,
it is possible to complete the circuit induced by the motion of the planet.}
\label{fig:move2}
\end{figure}

\begin{figure}
\includegraphics[scale=0.5]{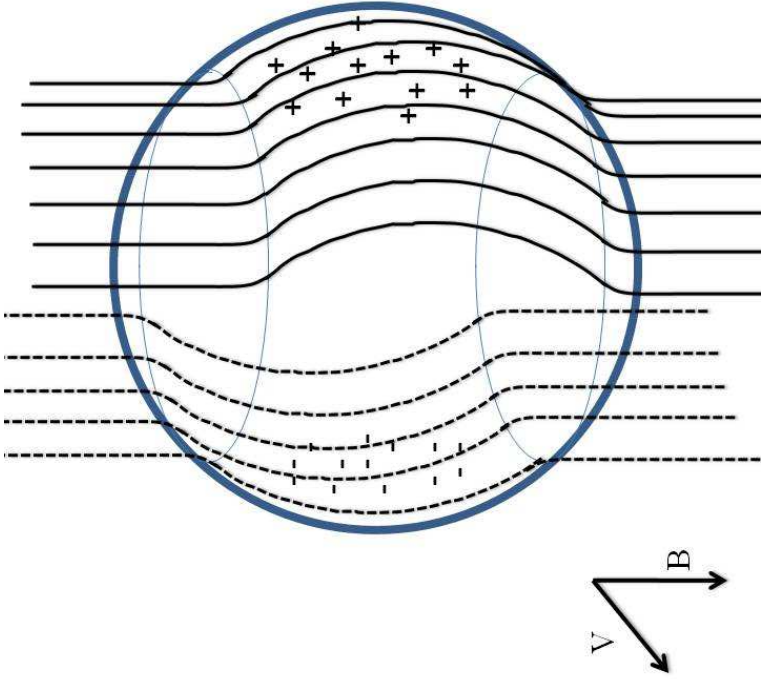}
\caption{Magnetic field lines in the case where the circuit 
is not closed. The planetary motion relative to the stellar magnetic 
field induces charge separation. Without any connection (at infinity)
between the separated charges, there is no current in the frame of 
the moving planet. In a infinitely conducting planet, the separated 
charges are concentrated near its surface. The magnitude of the 
induced field $\mathcal{E}_{p}$ is determined by Equation (\ref{eq:vcrossb}). 
In the stationary frame (centered on the host star), the separated 
surface charges carried by the planets generate two opposite currents 
as well as a finite $\nabla \wedge \mathcal{E}_{p}$ in the moving planet. 
Interior to the infinitely conducting moving planet, the 
induced field exactly cancels the unperturbed field as if there is
no diffusion of the stellar magnetic field into the planet. In 
the external region close to the moving planet, the induced field 
strongly perturbs the stellar magnetic field. The net field distortion
is symmetric around an infinitely conducting moving planet (analogous
to an invisic flow around a spherical object) such that there is
no net torque acting between the planet and the stellar field.
This symmetry would be broken and the drag on the planet would be 
finite if its conductivity is sufficiently low to permit significant 
slippage of the stellar magnetic field or if a complete circuit 
connecting two sides of the planet can be established on the surface 
of its host star.}
\label{fig:NoCircuit}
\end{figure}

In order to apply the model described above, one needs to verify that the 
time $t_{A}$ required for the Alfven waves to travel along the flux tube 
(to a depth $D_{pn}$ inside the star to 
be determined), and back to the planet is smaller than the time $t_{max}$ 
it takes the flux tube to slip ahead of the planet by more than its 
diameter. This condition ensures that a perturbation along a field line 
of the flux tube has the time to travel back and forth while the field 
line is still part of the flux tube that passes through the planet. 
Figures \ref{fig:move1} and \ref{fig:move2} illustrate this condition. 
Figure \ref{fig:NoCircuit} shows the field lines near the planet in 
the case where the condition is not met.  

\begin{figure}
\includegraphics[scale=0.5]{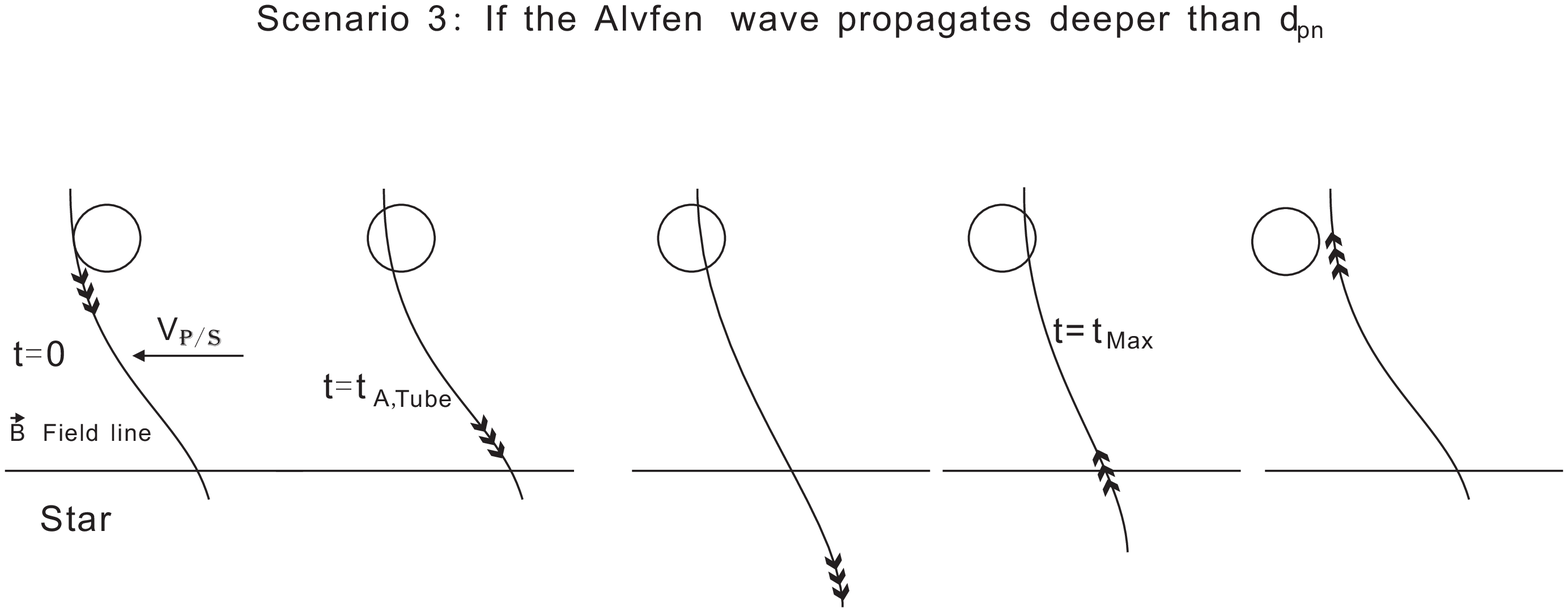}
\caption{Penetration depth of the unipolar induction circuit. 
The propagation of the induced disturbances from the planet to the stellar
surface is $t_{A, tube}$.  Below the stellar surface, gas density 
increases exponentially and the Alfven speed decreases accordingly.  
The penetration depth of the planet's induced field is determined by 
the requirement that the timescale for the Alfven waves to complete
the circuit equals to the field slippage timescale across the planet.
The net resistivity at the stellar surface is determined by the gas
across the flux tube in the form of parallel resistors.}
\label{fig:move3}
\end{figure}

In order to calculate $t_{A}$, we need to estimate the Alfven speeds 
along the flux tube between the planet and the star and in the stellar 
ionosphere (at the foot of the flux tube). Similarly, the calculation 
of $t_{max}$ requires the value of the conductivities (or resistance) 
of the different components of the circuit.  Indeed, the ratio of 
$\mathcal{R}_{p}$ (resistance of the planet) and $\mathcal{R}_{\ast}$ 
(resistance of the foot of the flux tube in the stellar atmosphere) 
determines the amount of relative slippage between the flux tube and 
the planet \citep{der70}.

In the limit where $\mathcal{R}_{p}$ is comparable to or larger than 
$\mathcal{R}_{\ast}$ (as in the night side of synchronously spinning
hot Jupiter, see Paper I), the flux tube would tend to slip through the 
planet. In this case, the flux tube would slip ahead of the planet 
by a distance $\sim 2R_{p}$ in a relatively short time $t_{A}$, and it 
might not be possible to maintain a closed circuit.  

In the most unfavorable case ($ \mathcal{R}_{p}/
\mathcal{R}_{\ast}=\infty$, 
i.e. the flux tube passes through the planet completely undisturbed), 
$t_{A}={2R_{p}} / {\upsilon_{p/s}}$ (where $\upsilon_{p/s}$ represents 
the speed of the planet in the frame rotating with the stellar 
magnetosphere). In the opposite extreme limit, $\mathcal{R}_{p}/
\mathcal{R}_{\ast} =0$ and $t_{A}=\infty$ such that the flux tube 
is completely anchored in the planet. Differential motion steadily
stretch the field lines until they reconnect. A more realistic situation 
falls somewhere between these two extreme limits, and the smaller 
$\mathcal{R}_{p}/\mathcal{R}_{\ast}$, the easier it is to satisfy 
the condition of validity.

\subsection{Qualitative Estimate of the Relative Slippage} 
A first qualitative criterion is given by the following argument.  We 
want to determine whether the magnetic flux tube slips on the planet 
or on the star.  In the absence of any companions, the magnetic 
field in the magnetosphere of a star rotates with the star. A close-in 
planet would tend to drag the stellar magnetic field lines that pass 
through it along with its motion.

Let us define $\omega_{p}$, $\omega_{\ast}$, and $\omega_{B}$ to be the
angular velocity of the planet, star, and magnetic field in an
absolute frame.  We then consider the relative motion between the
planet and the field lines, and between the field lines and the
star. We write $\Omega_{p}=\omega_{p}-\omega_{B}$ and
$\Omega_{\ast}=\omega_{B}-\omega_{\ast}$ and our goal is to estimate
the magnitude of $\Omega_{p}/\Omega_{\ast}$. In these notations, the
planet's speed relative to the magnetic field is then
$\upsilon_{p}=\Omega_{p}\ a$ ($a$ is the semi-major axis), and the
speed of the field lines (that pass through the planet) relative to
the star is $\upsilon_{\ast}=\Omega_{\ast} R_{\ast}$ ($R_{\ast}$ being
the stellar radius).

Considering the DC component of the field, we can write the complete
MHD induction equation (see Equation (\ref{MHD})) for the star and the planet.
\begin{eqnarray}
\frac{\partial \stackrel{\rightarrow}{\mathcal{B}}_{p}}{\partial
t} &=& \nabla \wedge (\stackrel{\rightarrow}{\upsilon}_{p}\wedge
\stackrel{\rightarrow}{\mathcal{B}}_{p}) +\nabla \wedge (\eta_{p}
\nabla \wedge\stackrel{\rightarrow}{\mathcal{B}}_{p}) \nonumber
\\ \frac{\partial \stackrel{\rightarrow}{\mathcal{B}}_{\ast}}{\partial
t}&=& \nabla \wedge (\stackrel{\rightarrow}{\upsilon}_{\ast}\wedge
\stackrel{\rightarrow}{\mathcal{B}}_{\ast}) +\nabla \wedge (\eta_{\ast}
\nabla \wedge\stackrel{\rightarrow}{\mathcal{B}}_{\ast}).
\end{eqnarray}
In a steady state, the first and second equations imply 
$\upsilon_{p}\approx {\eta_{p}} / {R_{p}}$ and 
$\upsilon_{\ast}\approx {\eta_{\ast}} / {R_{\ast}}$, 
respectively. Therefore, we obtain
\begin{equation}
\frac{\Omega_{p}}{\Omega_{\ast}}\approx
\frac{\sigma_{\ast}}{\sigma_{p}}\ \frac{R_{\ast}^{2}}{R_{p}a}.
\end{equation}

In the context of Io-Jupiter interaction ${R_{\ast}^{2}} /
{R_{p}a}\approx 7$. Since the electrical conductivity on Io is 
estimated to be much larger than that on Jupiter, the flux 
tube which passes through Io moves with Io and
drags its foot on the surface of Jupiter \citep{gold69}.
For a Jupiter-mass planet orbiting a young T-Tauri star (with 
radius twice that of the sun) at 0.04 AU (3 day-period), 
${R_{\ast}^{2}}/{R_{p}a}\approx 4$, and the anchorage of the flux tube
which passes through the hot Jupiters is thus determined by the ratio of
the diffusivity through the planet to that through its host star. For 
a super-Earth with radius twice that of the earth at 0.04 AU, 
${R_{\ast}^{2}}/{R_{p}a}\approx 20$.

\subsection{Analytical Expression of the Time Constraint, $t_{max}$}

The magnitude of $t_{max}$ is the time it takes for the field lines in
the flux tube to slip pass through the planet by a distance equal 
to the planetary diameter. We consider a planet with an electrical 
conductivity $\sigma_{p}$ moving relative to a magnetic field at speed 
$\upsilon_{p/s}=(\omega_{p}
-\omega_{\ast})a$. If the planet does not drag the field at all 
(for example if $\sigma_{p}=0$), the field lines would move 
relative to the planet with a linear speed $\upsilon_{p/s}$. Thus, 
the minimum value for $t_{max}$ is $t_{max}={2R_{p}}/
{\upsilon_{p/s}}$. On the other hand, if the field lines are 
perfectly anchored in the planet (for example when $\sigma_{p}
=\infty$) then $t_{max}=\infty$. The induced field lines would
wrap around the host stars with the planet's synodic orbit (i.e. 
its motion relative to the stellar spin).  

Based on extrapolation from analogous considerations (Aly 1985, 
Aly \& Kuijpers, van Ballegooijen 1994), we hypothesize that magnetic 
reconnection may occur when the azimuthal component of the 
induced (and ``dragged'') magnetic field well outside the planet 
becomes comparable to the unperturbed stellar dipole field.  
We assume the time growth timescale for the azimuthal component 
of the field to the period $t_{syn} = 2 \pi /(\omega_{p} -
\omega_{\ast})$ of the planet's synodic period. If it occurs, 
reconnection would lead to short-circuit and a burst of intense 
ohmic dissipation in the planet. We discuss the possibility of magnetic 
reconnection again in section (\ref{sec:approximations}). In the limit 
of infinite $t_{max}$, the relative motion between the planet and the 
stellar field would restore the electric field and re-establish the 
circuit on the timescale of $t_{A}$ with a reduced effective
conductivity (or equivalently an enhanced resistivity and magnetic
diffusivity). We shall consider elsewhere 
the possibility of such an episodic electrodynamic process.

In general, the conductivity $\sigma_{p}$ would fall between 
$0$ and $\infty$ and the field is dragged without being completely 
anchored on the planet. The induced field lines are distorted and  
partially wrapped around the host star. In the limit that $t_{A} 
< t_{max} < t_{syn}$, it is possible to complete a steady circuit 
of unipolar induction without reconnection.  

The same argument also holds for the star with electrical 
conductivity $\sigma_{\ast}$ dragging its own field lines so that 
the intensity of the slippage between the field lines and the planet 
depends on both the conductivity of the planet and star. Modeling 
the interaction between Io and Jupiter, Goldreich \& Lynden-Bell (1969) 
took into account $\sigma_{Io}$ and $\sigma_{Jupiter}$. Dermott (1970)
included the contribution of the flux tube's conductivity in the 
expression of the slippage of the field relative to both Io and Jupiter. 

In the present context, the linear speed of the slippage between 
the flux tube and the planet is
\begin{equation}
\upsilon_{slip}=\frac{\upsilon_{p/s}}{1+(\mathcal{R}_{\ast} +
\mathcal{R}_{tube})/(\mathcal{R}_{tube}+\mathcal{R}_{p})}.
\end{equation}
(In the above expression, we use $\mathcal{R}_{p}$ instead of 
$2 \mathcal{R}_{p}$ as in Dermott). 

In a complete circuit, the maximum time available for the 
Alfven waves to propagate from the planet to the star and return 
to the planet is 
\begin{equation}
t_{max}=\frac{2R_{p}}{\upsilon_{slip}}=2R_{p} \frac{(1+w)}
{(\omega_{p}-\omega_{\ast})a}
\label{tmax}
\end{equation}
where $w={(\mathcal{R}_{\ast}+\mathcal{R}_{tube})}/
{(\mathcal{R}_{p}+\mathcal{R}_{tube})}$. It is common to 
neglect $\mathcal{R}_{tube}$.

\section{Resistance and Alfven speed along the circuit}

We derived the analytical expressions of the total intensity 
(Section 3.4) and the time available for the Alfven waves to travel 
back and forth between the planet and the foot of the flux tube 
(Section 4.2). In order to determine their numerical values, we 
calculate in this section 1) the planet's integrated resistance 
$\mathcal{R}_{p}$, 2) the resistance $\mathcal{R}_{tube}$ and 
Alfven speed $\upsilon_{A,tube}$ along the flux tube, 3) and 
the resistance across the foot of the flux tube just below 
the star's surface $\mathcal{R}_{\ast}$ (perpendicular to the 
magnetic field and parallel to the electric potential gradient) 
and the Alfven speed along the foot of the flux tube 
$\upsilon_{A,tube}$. Figures \ref{fig:move1} and \ref{fig:move2} 
provide a summary of the condition of validity.

\subsection{Resistance in the Planet $\mathcal{R}_{p}$}
The electrical conductivity profile of a super-Earth is 
unclear. We thus first discuss the (better characterized) 
conductivity profile of present day Earth. Lorrain et al. 
(2006) estimated the electric conductivity of the 
present-day Earth mantle to range between $10^{-2}\ 
{\rm ohm}^{-1}\ {\rm m}^{-1}$ and $10^{3}\ {\rm ohm}^{-1} 
\ {\rm m}^{-1}$ and $10^{5}\ {\rm ohm}^{-1} \ {\rm m}^{-1}$ 
for the inner core (also see Stevenson, 2003). Merrill et al. 
(1996, pp. 273-277) similarly argues for an electrical 
conductivity of about 5-8 $\times 10^{5}\ {\rm ohm}^{-1} \ 
{\rm m}^{-1}$ for the core of the Earth, and between 3 and 
100 ${\rm ohm}^{-1} \ {\rm m}^{-1}$ for the lower mantle. 
For the upper mantle (and crust), Obiekezie \& Okeke (2010) 
calculate an electrical conductivity increasing from the 
surface (about $3 \times 10^{-2}\ 
{\rm ohm}^{-1}\ {\rm m}^{-1}$) to $10^{-1}\ {\rm ohm}^{-1}\ 
{\rm m}^{-1}$ at around 500km. The electrical conductivity 
of the Earth is therefore minimal and between $3 \times 
10^{-2}\ $ and $10^{-1} {\rm ohm}^{-1}\ {\rm m}^{-1}$) 
for a few hundred kilometers and then increase with depth. 

In the present application, we are primarily interested in 
the interaction between super-Earths and their host stars
when they are relatively young (up to a few $10^7$ yr).  During
this stage, the stellar magnetic field is intense and the 
close-in super-Earths may be intensely heated by giant impacts,
tidal and ohmic dissipation. Super-Earths with a molten crust
are likely to have higher conductivities than the present-date
terrestrial planets (for example, Rikitake (1966) expressed 
the conductivity of rocks and metals on the Earth as a sum 
of $exp(-E_{i}/kT)$, and Umemoto et al. (2006) estimated a 
conductivity at the core of a super-Earth and hot-Jupiter 
to be around $10^{6}\ {\rm ohm}^{-1} \ {\rm m}^{-1}$). The 
stellar radiation alone would raise the super-Earth surface 
temperature to about 1500K. Ohmic dissipation inside the 
planet may provide an additional source of thermal energy
(see Section \ref{sec:ohmplanet}). In a thermal equilibrium, the 
planet's surface temperature may sometimes exceed 2000K, 
at which silicate melts and raise the electric conductivity 
to around 10 ${\rm ohm}^{-1} \ {\rm m}^{-1}$ (for 1400K, 
Waff \& Weill, 1975).   

It is therefore reasonable to assume that the electrical 
conductivity of a super-Earth is most likely higher but at 
least that of the Earth. The electrical conductivity in a 
super earth would thus be several order of magnitude higher 
than $1 \ {\rm ohm}^{-1}\ {\rm m}^{-1}$ in the core and 
lower mantle and arguably also in the upper mantle. Besides, 
a conductivity of 0.1-$1 \ {\rm ohm}^{-1}\ {\rm m}^{-1}$ (i.e. 
10 times lower than the value we use) in an area spanning 10 
\% of the planet (roughly the thickness of the upper mantle) 
would at most double total the resistance. 

In addition, for a super-Earth, the characteristic speed (for 
example the planet linear speed in a frame co-rotating with the 
star) is much faster that for the field to diffuse across it,
i.e. $\upsilon \gg \frac{\eta}{L} = \frac{1}{\mu_{0}\sigma L}$ 
where $L$ is the characteristic length. Therefore, we can 
neglect the diffusion (second term on the right hand side of 
the MHD equation (Equation (\ref{MHD})) compared to the induction (first term).

The integrated resistance in the geometry of the planet is 
\begin{equation} 
\mathcal{R}_{p}=\frac{1}{S} \frac{L}{\sigma} = \frac{1}{R_{p}\sigma_{p}}
\end{equation}
where $S$ is the cross section. Depending of the geometry of the current 
inside the planet, the formula could have multiplicative factors, 
usually of order unity. 
 
With these approximations, we find
\begin{equation}
\mathcal{R}_{p} = \frac{1}{R_{p}\sigma} \ll 7 \times 10^{-8} {\rm Ohm}
\label{resistance planet}
\end{equation}
The value we use in our fiducial calculation is $\mathcal{R}_{p}=7 
\times 10^{-8} {\rm Ohm}$. 
If our assumption that the conductivity in a super-Earth is higher than
that in the present-day Earth is inappropriate, this resistance could be
higher.  But it may also be much lower if the super-Earth has a substantial
atmosphere which is extensively photo-ionized or a fully molten core
where the alkali metals are partially ionized.

\subsection{Resistance and Alfven Speed Along the Flux Tube}

Electrodynamics along the flux tube determines the propagation of 
the induced electric field between the planet and its host star.
The total resistance along the flux tube determines changes in the
electric potential at the foot of the flux tube. The Alfven speed
determines the propagation speed of the disturbance.

\subsubsection{Total Resistance Along the Flux Tube $\mathcal{R}_{tube}$}

The resistance of the flux tube is also difficult to estimate accurately. 
Goldreich \& Lynden-Bell (1969) simply assumed the electric conductivity to 
be infinite along the magnetic field lines and did not include 
$\mathcal{R}_{tube}$ in their equations. Dermott included 
$\mathcal{R}_{tube}$ in the equations but, during numerical applications, 
assumed it to be negligible in front of the resistance of the satellite Io. 
In all previous investigations, conductivity across the field lines in the
tenuous region between the planet and the star  is assumed to be negligible.

We provide here an estimate of the order of magnitude of the resistance of 
the flux tube. If we assume the plasma between the star and the planet to 
be fully ionized, the electric conductivity along (parallel) to the magnetic 
field line would 
\begin{equation}
\sigma_{0} = \frac{n_{e}e^{2}}{m_{e}\nu_{e}}
\label{eq:sigma0}
\end{equation}
where $n_{e}$ is the volumic number of free electrons, $e$ and $m_{e}$ are
the charge and mass of the electron, and $\nu_{e}$ the collisional frequency 
of the electrons with electrons and ions/protons (we assume these two 
collisional frequency to be the same). 

We take $\nu_{e}=\frac{n_{e}e^{4}}{16\pi \epsilon_{0}^{2} m_{e}^{2} 
<\upsilon_{e}>^{3}}$ with the electron thermal speed $<\upsilon_{e}> 
= \sqrt{\frac{2k_{B}T}{m_{e}}}$. We thus get $\nu_{e}=n_{e} 1.2 \times 
10^{-6} T^{-3/2}$  and $\sigma_{0}=\frac{e^{2}T^{3/2}}{1.2 \times 
10^{-6} m_{e}}$. In our model, we consider a star with a surface 
temperature $T_\ast =4000$K, and a planet with the equilibrium 
temperature $T_p \sim 1,500$K for the planet with an $a =$ 0.04 AU. 
For an average temperature of 2000K between the star and the planet, we 
find  $\sigma_{0} \simeq 2000\  {\rm Ohm}^{-1}\ {\rm m}^{-1}$ so that 
\begin{equation}
\mathcal{R}_{tube}=\frac{L}{\sigma_{0}S}=\frac{a}{(\pi R_{p})(R_{p} f _{tube})
\sigma_{0}}=\frac{4.6 \times 10^{-9}}{f_{tube}} {\rm Ohm}
\end{equation}
where $f_{tube}$ is between $0$ and $1$ such that $R_{p}f_{tube}$ 
is equal to the thickness of the volumic current that flows along 
the field lines. This resistance is usually negligible compared to 
the other resistances involved in the model (especially that of the 
star), except if the volumic currents are confined in an extremely 
thin layer at the surface of the flux tube. Therefore, we neglect
the potential difference, along each field line in the flux tube
between the surfaces of the planet and star.

\subsubsection{Travel Time Along the Flux Tube Between the Planet 
and (the Top of) the Stellar Surface}

We assume that the plasma between the star and the planet is fully 
ionized and estimate $\upsilon_{A,tube}$ under various different 
situations.

1) We first consider the epoch shortly after the super-Earth has
migrated to the stellar proximity through planet-disk tidal 
interaction.  In opaque inner regions of their natal disks, 
super-Earths' type I migration generally stalls at a radius $r$
where the $\Sigma_d$ has a positive radial gradient with a scale 
height $\Delta r = \Sigma_d / (\partial \Sigma_d / \partial r)$ 
which is a fraction ($\sim 0.1-0.2$) of $r$ \citep{mas06}.
Special locations include narrow transition regions between 
active inner region and dead zone as well as outer edge of 
magnetospheric cavity \citep{kret09}.  
  
We consider a super-Earth is embedded in a disk with an effective 
thickness $H_d \simeq c_s / \omega_p \sim 0.01-0.1 r$ and a steady 
state mass transfer rate ${\dot M}_d = 2 \pi \Sigma_d U_a r$ 
throughout the disk where $\Sigma_d$ and $c_s$ are the surface 
density and sound speed of the gas.  Using an {\it ad hoc} $\alpha$ 
prescription for the effective viscosity $\nu = \alpha \omega_p H_d^2$, 
the radial velocity of the disk gas is $U_d \simeq - 3 \nu / 2 r =
-3 \alpha H_d^2 \omega_p / 2 r$ and the characteristic density at the 
disk mid plane is
\begin{equation}
\rho_d \simeq \Sigma_d / 2 H_d ={\dot M_d \over 6  \pi \alpha 
\Omega H_d^3} 
\end{equation}
where $\alpha$ is the turbulent transport efficiency factor and may 
have an effective magnitude $\sim 10^{-2} - 10^{-3}$ (Hartmann et al. 1998).
In untruncated protostellar disks around classical T Tauri with ${\dot M}_d 
\sim 10^{-7} - 10^{-8} M_\odot$ yr$^{-1}$, $\rho_d \sim 10^{-6}-10^{-7}$ 
kg cm$^{-3}$ at the edge of the magnetospheric cavity $r \sim 0.04$ AU. 
The corresponding Alfven speed is
\begin{equation}
\upsilon_{A,tube} = \frac{m}{a^{2}\sqrt{\mu_0 \rho_d}}
\sim 10^5 {\rm m} {\rm s}^{-1}.
\end{equation}
Near the disk inner edge, the characteristic wave propagation timescale 
$t_{tube} \sim \Delta r / \upsilon_{A,tube} \sim 10^4$s may be 
too long to maintain a circuit. Note that if the super-Earth is stalled
near the transition region between active and dead zones, $t_{tube}$
would be longer not only because this region is further away from the 
host star, but also because $\Sigma_d$ interior to it does not vanish.

However, around stars with ages larger than $10^7$ yr, ${\dot M}_{d}$ 
may decline below that found around T Tauri stars 
and $t_{tube}$ can be reduced substantially. If the observed weak 
(or absences of) NIR excess around young stellar objects 
\citep{sic06} in clusters with age of $\sim 10$ Myr is due to the 
depletion of inner holes in both gas and dust, $\Sigma_d$ and 
hence $\rho_d$ would be substantially smaller than the values 
estimated above.  Thus in the post T Tauri phase, the Alfven speed 
around the host stars of super-Earths wound increase to sufficiently large
values to enable the circuit to be closed, especially if we take 
into account the resistances in the calculation of the speed of 
slippage through the planet.
 
2) Density around the flux tube does not decline indefinitely.  Even 
after the disk is completely depleted or truncated in the proximity of 
the planet's orbit, the planet may be surrounded by a spherically
symmetric component of stellar outflow with a speed $\upsilon_{flow}$ 
and a mass flux.
In this case, the volumic mass distribution is
\begin{equation}
\varrho_w(r)=\frac{\stackrel{\centerdot}{M}_w}{4\pi r^{2}
\upsilon_{flow}}
\end{equation}
The Alfven speed between the planet and the star at radius r (with the
origin at the center of the star) is thus
\begin{equation}
\upsilon_{A,tube} (r) = \frac{m}{r^{2}}\sqrt{\frac{\mu_0 \upsilon_{flow}}
{4 \pi \stackrel{\centerdot}{M}}}.
\end{equation}
Using $\upsilon_{flow}=100$ km s$^{-1}$, the numerical applications give 
$\upsilon_{A,tube}(R_{\ast}) \simeq 10^{8}$ $ms^{-1}$ and $\upsilon_{A,tube}
(a) \simeq 6 \times 10^{6}$ $ms^{-1}$. The
time it takes the Alfven waves to travel down the flux tube is
\begin{equation}
t_{tube}=\int dr\ \frac{r^{2}}{m}\sqrt{\frac{4 \pi
\stackrel{\centerdot}{M}}{\mu_0 \upsilon_{flow}}} =
\frac{a^{3}}{3m}\sqrt{\frac{4 \pi \stackrel{\centerdot}{M}}
{\mu_0 \upsilon_{flow}}}\left[ 1-\left(\frac{R_{\ast}}{a}\right)^{3}\right]
\label{ttube}
\end{equation}
where the integral being for r varying from the surface of the star to the
planet. If one takes $\stackrel{\centerdot}{M}=10^{-10} M_{\odot}$ yr$^{-1}$ 
and $\upsilon_{flow}=100$ km s$^{-1}$, the numerical application then gives 
$t_{A,tube} \simeq 300$s. The magnitude of $t_{A,tube}$ would be smaller 
for winds with faster speeds or lower mass loss rate.  

\subsection{Resistance in the Star Across the Foot of the Flux Tube and 
Alfven Speed Along the Magnetic Field in the Star at the Foot of the Flux Tube}
\label{sec:sigmastar}

The resistance of the foot of the flux tube determines the total 
intensity in the circuit, and most of the travel time of the Alfven 
waves occur at the foot of the flux tube. 

\subsubsection{Temperature and Pressure of the Stellar Outer Layer in 
an Isothermal Approximation}

For the outer layer of the star, we adopt an isothermal approximation 
and assume a spherical symmetry. The pressure and temperature $P(r)$ 
and T are then given by,
\begin{eqnarray}
\label{P}
T(r) &=& T_\ast \\
P(r) &=& P(R_{\ast})\ exp\left[\frac{GM_{\ast} \mu}{R_g T(r) R_{\ast}} 
\left(\frac{R_{\ast}}{r} - 1 \right) \right] \\
P(R_{\ast}) &=& \frac{2}{3}\frac{g_{s}}{\kappa}  
\end{eqnarray}
where $\kappa$ and $\mu$ are the opacity and molecular weight at the 
photosphere and $R_{g}={\mathcal{N}_{A}k_{B}}/{\mathcal{M}_{H}} 
\simeq 8.3 \times 10^{3}$ in SI units, with $\mathcal{N}_{A}$ the 
Avogadro number, $k_{B}$ the Boltzmann constant, and $\mathcal{M}_{H}$ 
the molar mass of the hydrogen atom. The volumic mass can also be calculated 
using the ideal gas equation of state. For the models 
presented here, we neglect any change in $T(r)$ and $P(r)$ due to the 
local ohmic heating at the foot of the flux tube.  Discussions in
Section \ref{sec:ohmicfoot} shows the possible existence of a hot spot at the 
foot of the flux tube.  Self consistent treatment of a potential 
feedback effect will be analyzed elsewhere.  

\subsubsection{Conductivity in the Stellar Atmosphere}

The details of the derivation of the conductivity in the stellar atmosphere 
(foot of the flux tube) are given in Appendix A. In Figure \ref{fig:foot},
we show that, at the foot of the flux tube, current flow across the 
field lines, as a series of parallel circuits.  We calculate the effective
resistance and Alfven travel timescale to determine the depth of 
penetration.  We use the Saha's equation 
to derive the ionization fraction. Following Fejer (1965) we refer
$\sigma_{0}$ as the electric conductivity parallel to the magnetic field 
lines, and $\sigma_{p}=\frac{\sigma_{0}}{1+(\omega_{e}/\nu_{e})^{2}}$ 
the Pedersen conductivity parallel to the electric field. 
We define $\omega_{e} = e B / m_{e}$ to be the electron gyro-frequency,
$\nu_{e}$ to be the mean collision frequency of the electrons with the 
neutral gas (see Equations (A7) and (A8)) and $r_{=}$ to be the radius at which 
\begin{equation}
\nu_{e}(r_{=})=\omega_{e}(r_{=}).
\end{equation}
We find $r_{=} \approx 1.3962 \times 10^{9}$m (given with several significant figures
as an intermediate value in the series of numerical applications).
Since $\omega_{e}(r) < < \nu_{e}(r)$ at $r \leq r_{=}$ and $\omega_{e}(r) 
> \nu_{e}(r)$ at $r\geq r_{=}$, we write the Pedersen conductivity
\begin{eqnarray}
\sigma_{p}(r \leq r_{=}) &=& \sigma_{0}(r) \\
\sigma_{p}(r \geq r_{=}) &=& {\sigma_{0}(r)}{\left(\frac{\nu_{e}}
{\omega_{e}}\right)^{2}}.
\label{eq:sigmap}
\end{eqnarray}

In contrast to the region between the planet and its host star, gas in
the stellar atmosphere is partially ionized.  Substituting the appropriate
value for $\sigma_{0}$ from Equations (A4) and (A5) we find
\begin{eqnarray}
\label{sigmaP}
\sigma_{p}(r \leq r_{=}) &=& L_a \ exp\left(\frac{-E}{2k_{B}T(r)} \right)\ 
\frac{T(r)^{3/4}}{\sqrt{P(r)}} \\ 
\sigma_{p}(r \geq r_{=}) &=& \frac{L_a }{Q_a ^{2}}\ \frac{1}{m^{2}T(r)^{1/4}}\ 
exp\left(\frac{-E}{2kT(r)} \right) \ r^{6}\ (P(r))^{3/2} \\
\frac{L_a }{Q_a ^{2}} &=& 10^{-19}\ \left(\frac{\mu_{0}}{4\pi} \right)^{-2}\ 
\frac{\left(2\pi m_{e}\right)^{3/4}}{h^{3/2}}  k_{B}^{-1/4}\ 
\left(\frac{128 m_{e}}{9\pi} \right)^{1/2}
\end{eqnarray}
where the numerical values of the constants in SI units are 
$L_a =6.17 \times 10^{6}$, $Q_a=2.93 \times 10^{-4}$, and 
${L_a}/{Q_a^{2}}=7.2 \times 10^{13}$.

\subsubsection{Alfven Speed and Resistance}
\label{sec:depth}
In Equation (\ref{Rstar}), we showed that $\mathcal{R}_{\ast}$ is 
given by
\begin{equation}
\mathcal{R}_{\ast}=\frac{1}{2\Sigma s},
\end{equation}
where s has been previously defined as $cos \theta_{F}$. 
Here, we decompose $\Sigma$, the integral from $r_{pn}$ to $R_{\ast}$ 
of the electric conductivity, into two parts $\Sigma_{1}$ and $\Sigma_{2}$, 
respectively the integral of the electric conductivity $\sigma_{p}(z)$ from 
$r_{pn}$ to $r_{=}$ and from $r_{=}$ to $R_{\ast}$ such that

\begin{eqnarray}
\Sigma &=& \int_{r} \sigma_{p}(r) dr = \Sigma_{1} + \Sigma_{2} \\
\Sigma_{1} &=& \int_{r_{pn}}^{r_{=}} \sigma_{p}(r) dr = 
L_a\frac{T(r)^{3/4}}{\sqrt{P(R_{\ast})}} exp\left(-\frac{E}{2k_{B}T(r)}\right) 
\int_{r_{pn}}^{r_{=}} exp\left[-\frac{GM_{\ast} \mu}{2 R_g T(r) R_{\ast}}
\left(\frac{R_{\ast}}{r}-1\right)\right]dr \\
\Sigma_{2} &=& \int_{r_{=}}^{R_\ast} \sigma_{p}(r)dr = \frac{L_a}
{Q_a^{2}} exp\left(-\frac{E}{2k_{B}T(r)}\right) \frac{(P(R_{\ast}))^{3/2}} 
{m^{2}T(r)^{1/4}} \int_{r{=}}^{R_\ast} r^{6} exp\left[\frac{3}{2} 
\frac{GM_{\ast}\mu }{R_g T(r) R_{\ast}} \left(\frac{R_{\ast}}{r}-1\right)
\right] dr
\end{eqnarray}
where $y_{1}=(\frac{R_{p}}{s}) \left({R_{\ast}}/{a}\right)^{3/2} $ and $y_{2}
=2 R_{p} \left({R_{\ast}}/{a}\right)^{3/2} $ are respectively the length and width 
of the foot of the flux tube, $r_{=}$ is defined above, and the 
penetration radius $r_{pn}$ is to be determined below (see 
Figure \ref{fig:foot}). The numerical application gives $\Sigma_{2} 
\approx 1.5 \times 10^{4} {\rm ohm}^{-1}$, and the analytical expression of 
$\Sigma_{1}$ depends on $r_{pn}$. Nevertheless, in our fiducial model, the 
numerical value of $\Sigma_{1}$ does not depend significantly on $r_{pn}$ and 
we get $\Sigma_{1} \approx 4.3 \times 10^{4} {\rm ohm}^{-1}$.
These values lead to $\mathcal{R}_{\ast}=8.6 \times 10^{-6}$ ohm. 

At the foot of the flux tube on the surface of the star, the volumic
mass is $\varrho(r)={P(r)m_{p}}/{k_{B}T}$ and the expression for the 
ionization $x$ is given in the appendix A. Therefore, the Alfven
speed in the stellar atmosphere and the time $t_{A}$ is thus
\begin{eqnarray}
\upsilon_{A,\ast}(r) &=& \sqrt{\frac{\mathcal{B}^{2}k_{B}T(r)}
{\mu_{0}m_{p}P(r)x(r)}} \\
&=&
W\ \frac{m}{r^{3}T(r)^{1/8}} exp\left(\frac{E}{4k_{B}T(r)}\right)
\frac{1}{\left(P(R_{\ast})^{1/4}\right)} exp\left[\frac{GM_{\ast} \mu}
{4R_g T(r) R_{\ast}}\left(1-\frac{R_{\ast}}{r}\right)\right] \\
W &=& \frac{1}{2\pi} \sqrt{\frac{\mu_{0}}{m_{p}}} 
\frac{h^{3/4}}{(2\pi m_{e})^{3/8}} \frac{1}{k_{B}^{1/8}}. \\ 
\label{valf}
\end{eqnarray}

In SI units, $W=0.038$, $\mathcal{B}(r)={2cos 
\theta_{F} \mu_{0}m}/({4\pi R_{\ast}^{3}})$ (with $cos \theta_{F}\approx 1$, 
and the integral going from $r_{pn}$ to $R_{\ast}$ ($r_{pn}$ is the 
radius that determines the effective height $R_{\ast}-r_{pn}$ of the 
foot of the flux tube). Using the fiducial values for the parameters, 
we get $\upsilon_{A,\ast} \simeq 2.6 \times 10^{8} exp\left[-717.5 
\left(\frac{R_{s}}{r}-1 \right) \right]$. Clearly, $\upsilon_{A,\ast}$ 
decreases sharply from the surface toward the interior. $r_{pn}$ is 
thus the smallest radius that still enables the model to be valid 
(i.e. the deepest that a perturbation of the field line can penetrate 
inside the star and back to 
the planet in less than $t_{max}$ (see Figure \ref{fig:move3}).

The time it takes for the Alfven wave to travel from the 
surface of the star to the bottom of the flux tube is
\begin{equation}
t_{A,\ast} = \int_z 
\frac{dz}{\upsilon_{A,\ast}(z)}
\label{tA}
\end{equation}

We then equate the total travel time 
$2(t_{A,\ast} + t_{tube})$ (there is a coefficient ``2" since the 
wave needs to go from the planet to the star and back to the planet) 
defined in Equations (\ref{tA}) and (\ref{ttube}) with the total 
time available $t_{max}$ defined in Equation (\ref{tmax}) 
\begin{equation}
2(t_{A,\ast} + t_{tube})=t_{max} 
\label{time condition}
\end{equation}
and solve for $r_{pn}$. The numerical application 
gives $r_{pn} \approx 1.3718 \times 10^{9}$m for a fast rotating 
star and $r_{pn} \approx 1.3746 \times 10^{9}$m for a slow rotating 
star (given here with several significant digits simply 
as an intermediate value in the thread of numerical applications). 
Having determined $r_{pn}$, one could now calculate 
self-consistently $\Sigma_{1}$ and $\mathcal{R}_{\ast}$.

\section{Numerical Applications and Discussion}

The quantities we have determined above are applicable for the fiducial
model we adopted here.  A more general application (for host stars of 
different masses) will be presented elsewhere.

\subsection{Numerical Applications}
\label{sec:numerical values}
The numerical values of the intensity, ohmic dissipation, 
and torque in the star and the planet respectively for a 
slow rotating star (with a spin period 8 days) and a fast rotating 
star (with a spin period 0.8 days) are given below. The values below 
for U and I correspond to one of the two circuits (each circuit has 
the same value of U and I), and the values for $\mathcal{P}$ and 
$\mathcal{T}$ are for the entire planet and entire star (both circuits combined). \\
$U=6.7 \times 10^{9}$V and $2.8 \times 10^{10}$V (for slow and fast 
rotator respectively) \\
$I=7.8 \times 10^{14}$A and $3.2 \times 10^{15}$A\\
$\mathcal{P}_{\ast} = 10^{25}{\rm W}$ and $2 \times 
10^{26}{\rm W}$ \\
$\mathcal{P}_{p}=4 \times 10^{22}{\rm W}$ and $7 \times 
10^{23}{\rm W}$ \\
$\mid \mathcal{T}_{\ast}\mid=\mid \mathcal{T}_{p} 
\mid =6 \times 10^{29}$N m and 
$ 2 \times 10^{30}$N m.

\subsection{Ohmic Dissipation at the Foot of the Flux Tube on the Star 
and Hot-Spots}
\label{sec:ohmicfoot}
A super-Earth orbiting at $a=0.04$ AU from its host star induces 
an ohmic dissipation at the foot of the flux tube on the surface 
of the star of $5 \times 10^{24} {\rm W}$ (for a stellar spin 
period of 8 days) and $9 \times 10^{25} {\rm W}$ (for a spin
period of $0.8$ days, which is about 1 to $5\%$ of the stellar 
luminosity ($L_{\ast,total}=3.6 \times 10^{26} {\rm W}$). 
This effect would create an observable hot-spot at the surface 
of the star. The rate of energy dissipation per surface area 
at the foot of the flux tube would be about $2 \times 10^{10} 
{\rm W} \ {\rm m}^{-2}$ which is three orders of magnitude 
higher than the intrinsic radiative flux from the surface of a 
typical T Tauri star. 

Since $\Sigma_1 > \Sigma_2$, most of the dissipation occurs
in the region between $r_{pn}$ and $r_{=}$.  We note that the
density scale at the photosphere ($R_\ast$), $\delta r_\ast =
R_g T_\ast R_\ast^2 / G M_\ast \mu \sim 5 \times 10^5$m is much 
smaller than $R_\ast - r_{=} = 3.8 \times 10^6$m and $R_\ast - r_{pn}
= 1.55 \times 10^7$m.  When the dissipated energy emerges from 
the stellar photosphere, the actual area of the hot spot may 
be diffused to several times the area of the foot of the flux 
tube.  The corresponding temperature of the hot spot would be
$\sim 2-3$ that elsewhere on the stellar surface.  In our
model, we consider a star with a surface temperature $T_\ast =$4000K.
Ohmic dissipation at the foot the flux tube increases the local
ionization, conductivity, current, and torque. We will construct
a self consistent model in a follow-up analysis.  

\subsection{Ohmic Dissipation in the Planet and the Induced Mass Loss}
\label{sec:ohmplanet}
In Paper I, we considered the structural adjustment due to
ohmic dissipation.  In this paper, we have not yet considered 
the structure adjustment of super-Earth structure due to 
ohmic dissipation.  For the slow-rotator model, the rate of 
ohmic dissipation is six times that the planet receives from 
its host star's irradiation. In the absence of any structural 
adjustment, the super-Earth may attain a thermal equilibrium with an
effective blackbody temperature $T_p \simeq$2,300 K 
(and much higher for the fast-rotator model). With this 
temperature, planet's core crust would surrounded by an ocean and 
an extensive atmosphere where water and hydrogen molecules 
readily dissociate but their ionization fraction remains 
negligible. The local density scale height is $\delta r_p 
= \lambda R_p$ where
\begin{equation}
\lambda = R_g T_p R_p / G M_p \mu \sim 1 / 17 \mu.
\end{equation} 

The density scale height of the hydrogen atoms is much larger 
than that of all other elements including carbon, oxygen,
and silicates.  The mean free path for a hydrogen atom
to collide with an heavy element with a density $n_z$ is
$l_{H-Z} = 1/(n_z A)$ where $A \sim 10^{-19}$ m$^2$ is
a typical cross section.  Within $\delta r_{de} \sim 2-3$ 
hydrogen atoms' scale heights there are so few heavy-elemental 
atoms left, they essentially become thermally decoupled, 
i.e. $l_{H-Z} > \delta r_{de}$ for hydrogen atoms.  
We refer this location as the decoupling radius 
$r_{de} = R_p + \delta r_{de} \sim (1.1-1.2) R_p$ and
the local hydrogen density at $r_{de}$ as $\rho_{de}$).  
The local temperature $T_{de} = T(r_{de})$ is set by the blackbody 
temperature $T_p$ of the heavy elements in the planet's 
photosphere. The magnitude of $\rho_{de}=\rho(r_{de})$ for a rocky 
or icy planet can be estimated to be 
\begin{equation}
\rho_{de} = 3 f_H M_p / 4 \pi R_p^3 {\rm exp} - \delta_{de} \sim 
10 \ {\rm kg} \ {\rm  m}^{-3}
\end{equation} 
where typical fractional abundance of the hydrogen atoms $f_H \leq 0.1$.

Next, we consider the possibility of significant loss
of hydrogen atmosphere. Planetary outflow is usually 
analyzed in the limit that atmosphere is heated by 
stellar irradiation.  For the present configuration, simple estimates 
indicate that the hydrogen atmosphere is opaque to 
incident ionizing photons from the host star, i.e. 
they are mostly absorbed by hydrogen atoms in the upper
atmosphere.  Provided hydrogen atmosphere remains mostly 
atomic, most visual stellar photons would stream pass it 
and be absorbed by heavy elements near $R_p$. Transit 
light curves of such a super Earth in Ly ${\alpha}$ photons 
would be much deeper than that for visual photons, as in 
the case of HD 209458b, \citep{vid03}.
Thus, most regions of the atmosphere is not affected by
either ohmic dissipation or irradiation.  

The most important heat input is the ohmic dissipation 
which takes place at the base of the atmosphere.  If
we neglect energy deposition and loss in the atmosphere,
our problem would reduce to a simple spherical
Bondi (or Parker) solution.  In a steady state, the 
governing continuity and momentum equations at a 
location $r$ would reduce to
\begin{equation}
{\dot M}_H = 4 \pi \rho_H V_H r^2
\end{equation}
\begin{equation}
V_H {\partial V_H \over \partial r} =
- {c_s^2 \over \rho_H} {\partial \rho_H \over \partial r}
- {G M_p \over r^2} + {3 G M_\ast r \over a^3}
\end{equation}
where the last term represents
host star's tidal force \citep{dob07}
and $V_H$ is the radial velocity. Together they reduce to
\begin{equation} 
{(V_H^2 - c_s ^2)\over r} {\partial {\rm ln} V_H \over
\partial {\rm ln} r} = {2 c_s^2 \over r} - {G M_p \over
r^2} + {3 G M_\ast r \over a^3}.
\end{equation}
Transonic point (where $V_H = c_s$) occurs \citep{lub75, gu03}
near the Roche 
radius $r \simeq R_R = (M_p/3 M_\ast)^{1/3} a$.  
Interior to the transonic point, flow is subsonic.

In order to make further progress, we need to estimate the 
energy budget of the atmosphere. At the base of the atmosphere,
ohmic dissipation occurs primarily due to collision of 
charged (provided by the heavy elements) and neutral particles.
Most of the dissipated energy is emitted to space at $R_p$ as
blackbody radiation.  Below $r_{de}$, hydrogen atoms attain 
$T_p (\sim 2,300$K) through conduction as all other particles.  
Above $r_{de}$, hydrogen atoms attain different density 
distribution.

For computation simplicity, let us first consider an
isothermal equation of state.  For an
analytic approximation, we neglect the advection 
contribution to the momentum equation and obtain
\begin{equation}
\rho (r) \simeq \rho_{de} {\rm exp} (r_{de}/ \lambda r - 1 /\lambda). 
\label{eq:denprof}
\end{equation}
At large distances ($r > (2-3) r_{de} \sim 3 R_p$) but still 
well within $R_R (\sim 10^{10}$ cm $\sim 7 R_p$), hydrogen's 
density approaches to an isochoric limiting value 
$\rho_\infty \sim 4 \times 10^{-8} \rho_{de} \sim 10^{-7}$ 
kg m$^{-3}$. The mass loss rate at $R_R$ \citep{li10}
becomes
\begin{equation}
{\dot M}_{\rm hydro} \simeq 4 \pi R_R^2 \rho_\infty c_s \sim 
5 \times 10^{12} {\rm kg} {\rm s}^{-1}.
\label{eq:mdot}
\end{equation}
At this rate, the total hydrogen mass in the planet $f_H M_p$
is depleted in a few Myr.  

Note that hydrogen atoms contained within this region is 
negligible compared with $M_p$ as we have assumed in the 
momentum equation. In addition, the collisional mean free 
path between hydrogen atoms $l_{H-H} = m_H / (\rho A)$ 
is small compared the density scale height $\delta r_p$ 
and more importantly $R_R$. In this limits, it is 
more appropriate consider outflow in the hydrodynamic limit 
\citep{mur09}, as we have done above, 
rather than use the Jeans' escape formula \citep{lec06}.

If the planet's atmosphere is maintained above the 
recombination temperature so that it is primarily 
compose of hydrogen atoms, the main cooling process
would be the emission of Ly $\alpha$ photons at a rate 
\begin{equation}
\Lambda \simeq 7.5 \times 10^{-20} x n_H^2 {\rm exp} - (1.2 
\times 10^5 K / T) {\rm J} {\rm m} ^{-3} {\rm s^{-1}}.
\end{equation}
After integrating over the entire volume $\sim 4 \pi R_R^3 /3$,
the total energy loss rate is $L_{{\rm Ly} \alpha} 
\sim 10^{16} x$ Watt which is substantially below 
fraction of the dissipated energy flux ($f_H \mathcal{P}_{p}$) 
carried by the hydrogen atoms.
(In the above estimate, we use the asymptotic value of 
$\rho_\infty$ to estimate $\Lambda$.)  However, with the
magnitude of $\dot M_{\rm hydro}$ in Equation (\ref{eq:mdot}),
we find that a significant fraction of $f_H \mathcal{P}_{p}$
may be advected with the escape hydrogen gas.

Based on hydrogen atoms' ineffective absorption and emission 
rates, it is natural to contemplate the possibility that 
the planet's atmosphere expands adiabatically. In the limit
that ohmic dissipation provides the only source of heating
at its base, planet's atmosphere may be convectively
unstable.  Efficient convection also leads to constant entropy.

Using the conventional polytropic approximation (in which 
$P = K \rho ^\gamma$ and $\gamma = 1.4$) for an adiabatic 
hydrogen atmosphere, a stationary quasi hydrostatic 
solution can be constructed with
\begin{equation}
1 - \left( {\rho(r) \over \rho_{de}} \right)^{\gamma-1}
= \left( {\gamma - 1 \over \lambda} \right) 
\left( 1 - {R_{de} \over r} \right).
\end{equation}
The above equation implies that with an adiabatic equation 
of state, both density and temperature in the hydrogen 
atmosphere vanishes within $\delta r_{de} / (\gamma-1) 
\sim 2.5 \delta r_{de}$. Unless planet's photospheric radius 
can expand (see paper I) significantly, there would be no 
outflow, despite the intense ohmic dissipation below $R_p$, and 
all the thermal energy generated would efficiently radiated 
by atomic emission from heavy elements.  

However, planet's atmosphere may be heated to prevent 
its temperature from plummeting below that ($\sim 2000$K)
for hydrogen molecules to recombine. Rotational and 
vibrational bands of hydrogen molecules not only provide 
emission mechanisms but also opacity sources to absorb
the incident stellar irradiation and to diffuse thermal 
energy in the planet's atmosphere from its heated base
to its upper layers. In the mildly heated case, we 
anticipate the planet's hydrogen atmosphere to attain 
an equilibrium temperature so that the incident 
deposition of stellar photon energy would be balanced 
by planet's reprocessed luminosity.  (For our 
fiducial model, the equilibrium temperature is $T_p 
\sim 2000$K.)  In this limit, the loss of planet's 
hydrogen atmosphere relies more critically on the 
atmosphere ability to maintain a shallow temperature gradient 
than that to generate energy through ohmic dissipation.
This situation has already been analyzed in the context
of HD 209458b \citep{mur09}.

However, enhanced sources of energy may also expand the 
radius of planet's photosphere well beyond $R_p$.  This 
is a distinctive possibility for the fast-rotator model
in which case $\mathcal{P}_{p}$ is another 18 
times larger. This increase in the ohmic dissipation rate 
is due to the relatively large differential motion between
the planet and the magnetosphere of its host star.  If the
planet's photosphere remains at $R_p$, the enhanced energy 
source would increase $T_p$ by a factor of 2 which is 
comparable to the magnitude of $T_\ast$. At this temperature,
opacity due to H$^-$ process becomes significant. Planet's
envelope and photosphere may well expand, leading to a
possible runaway ohmic heating.  The above discussion 
clearly warrants further discussions and detailed 
treatments of radiation transfer in this type of super Earths
We shall carry out and present these analysis in a future paper.

The loss of planet's atmospheric hydrogen is likely to occur 
on a more rapid pace. It remains to be demonstrated that 
for the intense heating cases, how far up in the atmosphere
does thermal decoupling between hydrogen and heavy elements
occur.  If the planet's photosphere is well within $R_R$,
the density scale height of most other heavy elements such 
as carbon and oxygen above are sufficiently small that they
may be effectively retained near $R_p$. Oxygen atoms may 
combine with Mg, Fe, Ca, Na, Al, and Ti silicates to 
form high-density minerals such as enstatites, olivines, 
and pyroxenes. Planets composed mostly such substances
are expected to have compact sizes (Valencia et al. 2010).
Thus, it is likely that super Earths which 
migrated early to the proximity of their strongly 
magnetized host stars may attain relatively compact sizes
as in the case of COROT 7-b \citep{leg09} and planets around
Kepler 11. 

The rate of ohmic dissipation in short-period super Earths 
is likely to diminish as their host stars magnetic field
weakens with age. As their semimajor axis increases, planets 
which undergo outward migration around rapidly spinning 
host stars also encounter less intense stellar dipole field. 
Some residual oxygen atoms in the atmosphere may recombine 
to form oxygen molecules during the decline of the ohmic 
dissipation rate. Oxygen molecules are particularly 
important because they have been suggested as a bio-marker 
for the detection of life elsewhere in the Universe 
\citep{des02}.

\subsection{Discussion about Some Approximations}
\label{sec:approximations}
We presented here a preliminary model for the unipolar induction model.
Some approximations were made for computational convenience, and we 
briefly discuss here the validity of the approximations which have 
not yet been discussed in the paper. 

We only took into account hydrogen for the calculation of the 
conductivity/resistance of the star.  In a realistic model, especially
for low-mass stars, other elements may become important contributors of
the ionization fraction.  A more comprehensive study will be presented
elsewhere. In addition, we suggested that the foot of the flux tube 
at the stellar atmosphere would be significantly heated. However, we used 
T=4000K for the stellar surface temperature (usual T Tauri star). 
The feedback on the stellar temperature due to the circuit may be included in later models. 

\textit{For the evaluation of the planet's electrical resistivity}. 
The value we adopted ($7 \times 10^{-8}$ ohm) seems to be 
the most uncertain value in our calculation. This value of the 
resistance of the planet used here is likely to be a lower 
boundary for the mantle of the planet (a metallic core might 
have even higher conductivity). If the real resistance where 
to be lower, then 1) 
the time $t_{max}$ available for the Alfven waves to travel 
around the circuit would increase, which would result in a 
deeper foot of the flux tube and would also enable the model 
to hold for larger semi-major axes and 2) the ohmic dissipation 
in the planet would decrease. Nevertheless, an increase 
in the depth of the foot of the flux tube would not affect 
much the total resistance of the foot of the flux tube 
$\mathcal{R}_{\ast}$.

\textit{Induction at the foot of the flux tube}. 
Since the conductivity is very high along the magnetic field lines, 
the plasma in the star's magnetosphere rotates with the magnetic 
field lines. Therefore, the plasma contained in the flux tube also 
moves with the magnetic field lines as they are dragged along by the 
planet, and thus moves relative to the unperturbed magnetic field line. 
Therefore, just as the induction in the planet is due to the relative 
motion between the frame co-moving with the planet and the frame 
rotating with the magnetosphere, there can also be a magnetic 
induction in the plasma enclosed by the foot of the flux tube. The 
order of magnitude of this phenomenon will be at most comparable to 
the order of magnitude of the phenomenon presently described. Ferraro 
\& Plumpton (1966) provide a brief discussion of the problems raised 
by two good concentric conductors rotating in a magnetic field at 
different angular speeds. 

As mentioned earlier, there may be magnetic reconnection if the 
induced field dominate over the unperturbed stellar dipole field.  
The field lines also tend to wrap around the planet when the 
synodic period ($T_{synod}=\frac{2\pi}{\omega_{p}-\omega_{\ast}}$) 
is small compared to $t_{max}$, the time required for the field 
lines constituting the flux tube to 
move across the diameter of the planet. Using (\ref{tmax}), we find that 
$t_{max}=\frac{R_{p}}{\pi a}(1+w)T_{synod}$ with 
$w=\frac{(\mathcal{R}_{\ast}+\mathcal{R}_{tube})}
{(\mathcal{R}_{p}+\mathcal{R}_{tube})} \approx 
\frac{\mathcal{R}_{\ast}}{\mathcal{R}_{p}}$. 
For our parameters, this corresponds to $t_{max} \approx 
0.091 T_{synod}$ 
(using $\mathcal{R}_{p}=7 \times 10^{-8}$ ohm according to 
(\ref{resistance planet})
and $\mathcal{R}_{\ast}=8.6 \times 10^{-6}$ ohm according 
to section (\ref{sec:depth})). 

This also provides an upper limit on the ratio of resistances w
in order to stay with a model without reconnection.  
Indeed, magnetic reconnection may occur when $t_{max}$ is larger than a few $T_{synod}$, 
and one would thus arguably stay in the regime without frequent magnetic reconnection when 
\begin{equation}
1+w \leq K \frac{\pi a}{R_{p}} 
\label{upper limit}
\end{equation}
with K larger than 1, and w as defined above. 
Using $R_{p}=2R_{\oplus}$ and a= 0.04AU, we find 
$\frac{\pi a}{R_{p}} \approx 1300$. 
We have neglected the resistance of the 
flux tube $\mathcal{R}_{tube}$ in front of $\mathcal{R}_{p}$ and 
$\mathcal{R}_{\ast}$ but this approximation may break down in extreme cases. 
Nevertheless, in most cases, Equation (\ref{upper limit}) means that 
$\mathcal{R}_{\ast} / \mathcal{R}_{p}$ is smaller than 1300. 
The resistance of the planet depends mainly on its composition, and structure, 
which would adjust to the strong ohmic dissipation in its interior. 
The resistance of the foot of the flux tube in the stellar atmosphere would depend on 
the metallicity and the temperature, which would also adjust 
to the strong ohmic dissipation. 

Previously in the paper, we also discussed that the travel time of the Alfven wave can be at most 
$t_{max}$ (Equation (\ref{time condition})). $\mathcal{R}_{\ast}$, through its relationship 
with the variable depth of penetration, 
would self-consistently adjust depending on the parameters of the model. 
Indeed, larger $\mathcal{R}_{\ast}$ leads to larger w, then larger $t_{max}$, and 
thus larger depth of penetration since the Alfven waves 
have more time to travel between the planet and the star along the flux tube, into the stellar atmosphere at the foot of the flux tube, and back to the planet. 
Deeper depth of penetration then results in smaller $\mathcal{R}_{\ast}$ 
(equivalent resistance with resistances in parallel). 

The value $\mathcal{R}_{p}$ is less directly constrained by the model although 
it of course depends on the parameters chosen for the model. Nevertheless, changes in the value of 
$\mathcal{R}_{p}$ would result in adjustments in $\mathcal{R}_{\ast}$ through the mechanism mentioned just above. 

Goldreich and Lynden-Bell also interpret the torque calculated above 
(section \ref{sec:numerical values}) in terms of a 
toroidal magnetic stress due to a distortion of $B_\phi$ in the 
azimuthal direction, i.e. the direction of the motion.  
Neglecting the induced field in the $r-z$ (meridional) direction, 
they determined the longitude of the flux tube from the ratio
of the induced $B_\phi$ and the unperturbed stellar-dipole field.
They then determined, for the Jupiter-Io system, 
the forward-sweeping angle (or the backward-
sweeping angle in the case of a slowly rotating star) 
of each field line as it leaves the Io to be 13 deg.  
When a similar approach is
adopted in the present model, we find this angle may be close 
to 90 deg.  This large distortion is due to a strong torque
induced by the unipolar circuit with a relatively small 
$\mathcal{R}_{\ast}$ (and thus a large intensity).  
For such a large field distortion, 
Goldreich and Lynden-Bell suggested that the induction circuit may be
broken by field reconnection.  We shall further examine this
possibility elsewhere and determine whether it may significantly
weaken the effective torque.

\section{Summary and discussions}
\label{sec:summary}
With the advent of high-precision radial velocity and transit
surveys, we have entered an era of super-Earth discovery.
Although the detection probability (due to observational 
selection effects) decreases with planets' period, three times 
more planets are found with period between 3 and 10 days than 
between 1 and 3 days. We suggest that super-Earths' interaction with
the magnetosphere of their host stars may be one possible 
mechanism for this dichotomy. 

In this paper, we analyze the electrodynamics of super Earths 
orbiting in the proximity of strongly magnetized T Tauri stars.
We constructed a fiducial model in which the planet's orbital
frequency is not synchronized with the star's spin.  Their 
relative motion enables the planet to continually encounter
field lines which are locked on the star. As a good (but not
perfect) conductor, an emf is induced across the planet 
(along the semi-major axis). We estimate planet's 
conductivity and show that the 
stellar fields slip through the planet with a drift speed
considerably slower than its Keplerian speed.

We show that conductivity along the field line is likely to
be large and the perturbed potential (due to the induced
electric field) propagates along a flux tube away from the 
planet with an Alfven speed.  We show that for planets 
with period less than 3 or so days, the disturbance 
can reach the surface of the star and return before 
stellar fields have drifted through the planet.  

The foot of the flux tube is implanted to stellar surface.
As density increases with depth below the photosphere, 
the Alfven speed decreases. Penetration depth of the 
flux tube is determined by the condition that the timescale 
required for Alfven waves to complete a circuit
between the planet and its host star is comparable to that
for the stellar field to drift through the planet.  

Across the foot of the flux tube on the stellar surface,
the potential drop induces a current to flow across it.
We show that the resistance on the surface of the star
is larger than that in the planet.  Consequently, the 
intensity of the current is determined by the resistivity 
on the star. We quantitatively determine this resistivity, the 
associated current, ohmic dissipation rate, and torque
due to the Lorentz force. The ohmic dissipation in the star 
at the foot of the flux tube could also induce an observable 
hot-spot. 

The source of energy is the differential motion between
the planet and the magnetosphere of its host star.
The Lorentz force on the planet and its host star
leads to an evolution toward a state of synchronous rotation.  
Inside the corotation radius, planets tend to lose 
angular momentum and migrate inward and the opposite 
trend occurs outside the corotation radius.  Consequently
planets inside corotation migrate inward and those outside
corotation migrate outward.

For super-Earths with periods less than 3 days, the timescale 
for orbital evolution can be comparable or shorter 
than a few Myr (the timescale over which intense stellar
magnetic field is maintained).  The low abundance of
super-Earths with period less than 3 days may be due to 
their infant mortality.  

Due to their finite conductivity, ohmic dissipation also
occurs on within the super-Earths.  The heating rate
depends on planet's poorly determined resistivity.
Its magnitude can be comparable to or larger than that 
the planet received from the stellar irradiation.  The
intense rate of ohmic dissipation may cause water and 
hydrogen molecule to dissociated and hydrogen atoms 
to segregate from other heavy elements.  It is unclear 
whether a substantial fraction of the hydrogen atom may
escape though hydrodynamic outflows. As the field decay
with maturing stars, remaining excess oxygen atoms may 
either be incorporated in high density minerals or form 
oxygen molecules.  Either of these processes can lead to
consequences which may be observable in the near future.

There are several uncertainties which warrant further
investigation.  Conductivity in super-Earths and 
their host star need further study.  We have not yet 
apply these results to a wide range of stellar and 
planetary models.  The effect of feedback due to the
adjustment of planet's and star's heated atmosphere also need to 
be examine.  Perhaps the largest uncertainty 
is whether the intense induced field can lead to magnetic
reconnection and the breaking of the circuit.  Reconnection
would increase the effective magnetic diffusivity and
severely weakens the effective torque.  

In order to directly compare with
observations, we also need to consider a diverse range
of planetary orbits.  For example, this process may 
not work for planets with period longer than a few days.
Finally, it would be of interest to determine whether
the intense electromagnetic interaction between super
Earths and their host stars can be directly observed
in the radio-wave frequency range.  

Nevertheless, we show that electrodynamic
interaction is an important process for the
orbital and structure evolution of super-Earths
as well as hot Jupiters.  Along with many other
physical processes it introduces diversity
in the present-day configuration of extra solar
planetary systems.  

\appendix

\section{Electric conductivity at the foot of the flux tube}
The Saha's equation gives the ionization fraction of the hydrogen 
atom $x=\sqrt{\frac{K_{H}}{1+K_{H}}}$, where 
\begin{equation}
\label{KH}
K_{H}(r)=\frac{1}{P(r)}\frac{(2\pi m_{e})^{3/2}}{h^{3}} 
(kT)^{5/2}\ exp\left(-\frac{E}{kT}\right) 
\end{equation}.
where $P(r)$ is the pressure, $m_{e}$ the electron mass, $h$ 
Planck's constant, and $E$ the ionization energy of the hydrogen 
atom. $K_{H}$ is a function of r which, in the isothermal region, 
decreases as one moves from the surface of the star toward the 
interior. Since, in the situations considered in this paper, 
$K_{H}$ is small compared to unity at the surface of the star, 
we get the following expression for the ionization rate $x$ 
everywhere in the isothermal region
\begin{equation}
\label{Link KH and x}
x(r) \approx \sqrt{K_{H}(r)}.
\end{equation}
Using the formulas given by Fejer (1965), we calculate the electric 
conductivity profile in the stellar (isothermal) outer layer. The 
conductivity $\sigma_{0}(r)$, which determines the current parallel 
to the magnetic lines of force, is given by
\begin{equation}
\label{sigma0}
\sigma_{0}(r) = x(r)\ 10^{19} \frac{e^{2}}{m_{e}} \sqrt{\frac{9 
\pi m_{e}}{128 k_{B} T (r)}} \\
\end{equation}

Using (\ref{KH}) and (\ref{Link KH and x}), we obtain the following 
expression for $\sigma_{0}$
\begin{eqnarray}
\label{conductivity}
\sigma_{0}(r) &=& L_a \ exp\left(\frac{-E}{2kT (r)}\right)\ 
\frac{T(r)^{3/4}}{\sqrt{P(r)}} \\
L_a &=& 10^{19} \frac{e^{2}}{m_{e}} \sqrt{\frac{9 \pi 
m_{e}}{128 k_{B}}} \frac{\left(2 \pi m_{e}\right)^{3/4}}
{h^{3/2}} k_{B}^{5/4}
\end{eqnarray}
which decreases as one moves from the surface of the star 
toward the interior. The numerical value of the constant $L_a$ 
in SI units is $L_a \simeq 6.17 \times 10^{6}$

The (Pedersen) conductivity $\sigma_{p}(r)$, which determines the 
current parallel to the electric field, is given by 

\begin{equation}
\sigma_{p} = \frac{\sigma_{0}}{1 + \left(\frac{\omega_{e}}
{\nu_{e}}\right)^{2}}
\end{equation}
where $\omega_{e}$ (the gyro-frequency of the electron) and 
$\nu_{e}$ (in the limit of a gas with low ionization fraction, 
$\nu_{e}$, is related to the mean collisional frequencies of the 
electrons with molecules of the neutral gas, see Draine et al. 1983) 
are given by
\begin{eqnarray}
\label{omega and nu}
\omega_{e}(r) &=& \frac{e \mathcal{B}_{s}(r)}{m_{e}} \\
\nu_{e}(r)    &=& 10^{-19}\ n \left(\frac{128kT(r)}{9 \pi 
m_{e}}\right)^{1/2} = 10^{-19}\ P(r)\ \left(\frac{128}
{9\pi m_{e}kT(r)}\right)^{1/2}
\end{eqnarray}
with $n$ the number density of neutral particles. Since the ionization 
rate is small, $n$ is also the number density of particles which is 
equal to ${P}/{kT}$ for a perfect gas. Using (\ref{omega and nu}) 
and the expression of a dipole magnetic field $\mathcal{B}_{s}(r) = 
{\mu_{0}\ m}/{(4\pi r^{3})}$ (with $m$ being the stellar magnetic moment), 
we obtain
\begin{eqnarray}
\label{omega over nu}
\frac{\omega_{e}}{\nu_{e}}(r) &=& Q_a\ \frac{m\sqrt{T(r)}}{r^{3}P(r)}  \\
Q_a &=& \frac{e}{m_{e}} \frac{\mu_{0}}{4\pi} 10^{19} 
\left(\frac{9\pi m_{e} k_{B}}{128}\right)^{1/2}
\end{eqnarray}
where the numerical value of the constant $Q_a$ in SI units is $Q_a=
2.93 \times 10^{-4}$.

In order to compare $\frac{\omega_{e}}{\nu_{e}}(r)$ with unity, 
we define $r_{=}$ such that 
\begin{eqnarray}
\frac{\omega_{e}}{\nu_{e}}(r_{=}) = 1
\end{eqnarray}
or, equivalently,
\begin{equation}
P(r_{=}) = \frac{Q\ m\sqrt{T(r)}}{r^{3}_{=}}.
\end{equation} 
Using the numerical values for the star listed above, we deduce 
$r_{=}=1.3962 \times 10^{9}$ (note that in paper I, what we defined 
$r_{=}$ to be the transition between the isothermal and the polytropic 
region in the star, which is unrelated quantity defined here). 

Since $P(r)$ increases rapidly when $r$ decreases (from the stellar 
surface inward), one may distinguish two regimes by
\begin{eqnarray}
\sigma_{p}(r \leq r_{=}) &=& \sigma_{0}(r) \\
\sigma_{p}(r \geq r_{=}) &=& \frac{\sigma_{0}(r)}{\left(
\frac{\omega_{e}}{\nu_{e}}\right)^{2}}
\end{eqnarray}
i.e. 
\begin{eqnarray}
\label{sigmaP}
\sigma_{p}(r \leq r_{=}) &=& L_a\ exp\left(\frac{-E}{2k_{B}T(r)} \right)\ 
\frac{T(r)^{3/4}}{\sqrt{P(r)}} \\ 
\sigma_{p}(r \geq r_{=}) &=& \frac{L_a}{Q_a^{2}}\ 
\frac{1}{m^{2}T(r)^{1/4}}\ exp\left(\frac{-E}{2kT(r)} \right) 
\ r^{6}\ (P(r))^{3/2} \\
\frac{L_a}{Q_a^{2}} &=& 10^{-19}\ \left(\frac{\mu_{0}}{4\pi} 
\right)^{-2}\ \frac{\left(2\pi m_{e}\right)^{3/4}}{h^{3/2}}  
k_{B}^{-1/4}\ \left(\frac{128 m_{e}}{9\pi} \right)^{1/2}
\end{eqnarray}
where the numerical value of the constant in SI units is 
${L_a}/{Q_a^{2}}=7.2 \times 10^{13}$.

\section{Another estimate of the Alfven speed along the flux tube}
The volumic current passing through the tube is $\mathcal{J} 
= e\upsilon_{e} {f\varrho}/ {m_{p}}$. In this expression, 
we adopt the current propagation $\upsilon_{e}$ to be the 
thermal speed of the electrons $m_{e} \upsilon_{e}^{2}/2=k_{B}T$ and we 
define $f$ to be the fraction of gas particles that are ionized. The
total flux $\mathcal{J}$ is the total intensity I divided by the cross
section of the flux tube through which the current passes. It is a
fraction $g_{2}$ of the cross section $\pi R_{p}^{2}$ of the planet. With
these notations, we get $\mathcal{J}={I}/{g_{2}\pi R_{p}^{2}}$.

We thus obtain the following Alfven speed and $t_{A,tube}$
\begin{eqnarray}
\upsilon_{A,tube} &=& \frac{\mu_{0}m}{4\pi}
\sqrt{\frac{1}{\mu_{0}}\frac{\pi R_{p}^{2}}
{I}\frac{e}{m_{p}}}\left(\frac{2k_{B}T}{m_{e}}\right)^{1/4}
\frac{\sqrt{g_{2}}}{r^{3}} \\
t_{A,tube} &=& \frac{2}{m} \sqrt{\frac{\pi I m_{p}}{\mu_{0} 
R_{p}^{2} e}} \left(\frac{m_{e}}{2k_{B}T}\right)^{1/4} a^{4}
\left[1-\left(\frac{R_{\ast}}{a}\right)^{4}\right] \frac{1}{\sqrt{fg_{2}}}
\end{eqnarray}
which gives, for $fg_{2}=1$, $t_{A,tube} \simeq 450$s.

\acknowledgments

We thank A. Cumming, F. de Colle, S.F. Dong, G. Ogilvie, G. Glatzmaier, and Q. Williams 
for useful discussions.  This work is supported by NASA 
(NNX07A-L13G, NNX07AI88G, NNX08AL41G, and NNX08AM84G), and NSF(AST-0908807).

\end{document}